\documentclass{elsarticle}

\usepackage{amssymb,amsmath,graphicx,url}
\usepackage[utf8]{inputenc}
\usepackage{listings}
\usepackage{tabularx}
\usepackage{natbib}
\usepackage{lscape}
\usepackage{graphicx}

\begin{document}

\title{Automatic Construction of Multi-faceted User Profiles using Text Clustering and its Application to Expert Recommendation and Filtering Problems}

\author{Luis M. de Campos}
\ead{lci@decsai.ugr.es}
\author{Juan M. Fernández-Luna \corref{cor1}}
\ead{jmfluna@decsai.ugr.es}
\author{Juan F. Huete}
\ead{jhg@decsai.ugr.es}
\author{Luis Redondo-Exp\'osito}
\ead{luisre@decsai.ugr.es}

\cortext[cor1]{Corresponding author. Tel.: +34958240804;}
\address{Departamento de Ciencias de la Computaci\'on e Inteligencia Artificial, ETSI Inform\'atica y de Telecomunicaci\'on, CITIC-UGR, Universidad de Granada, 18071, Granada, Spain}

\begin{abstract}

In the information age we are living in today, not only are we interested in accessing multimedia objects such as documents, videos, etc. but also in searching for professional experts, people or celebrities, possibly for professional needs or just for fun. Information access systems need to be able to extract and exploit various sources of information (usually in text format) about such individuals, and to represent them in a suitable way usually in the form of a profile. In this article, we tackle the problems of profile-based expert recommendation and document filtering from a machine learning perspective by clustering expert textual sources to build profiles and capture the different hidden topics in which the experts are interested. The experts will then be represented by means of multi-faceted profiles. Our experiments show that this is a valid technique to improve the performance of expert finding and document filtering.

\end{abstract}

\begin{keyword}
Clustering \sep Content-based Recommendation \sep Expert Finding \sep Filtering \sep User Profiling
\end{keyword}

\maketitle

\section{Introduction} \label{section:intro}

The content of the world wide web is incredibly wide and varied and so one common search task is to look for people that can help us with a particular problem. For example, we might search for a doctor to treat a specific illness, a builder to repair a leaking roof, or a politician to discuss a local problem with so that solutions may be found. This type of information search is set in the broader field of expert finding \citep{BFMS12} whereby users find experts in a given area. For this task to be successful, it is necessary for experts to be represented in some way in the retrieval system. The most specialized and accurate way is to consider experts' profiles as these store the most representative keywords to define their areas of expertise. These profiles would be built by considering the documents that best represent the expert: for example, for scientists, this would be their journal or conference publications; for writers, their published books; for programmers, the source codes they have written; for lawyers, the court cases they have worked on; and for politicians, their interventions in parliamentary sessions.

With all of these documents, a system could automatically build expert profiles by selecting the best keywords for the expert's fields of expertise. This source of information would then be used by the expert finding system to match the user's information requirements represented in the form of a query. There are basically two main problems related to finding relevant people where profiles are used:

\begin{itemize}
\item Given a group of experts or professionals, the problem consists in returning the most suitable ones that could fit a need expressed by a user (usually in the form of a short query). In this case, only the top-ranked ones will be recommended. This is considered to be an expert-finding problem or, more broadly speaking, content-based recommendation \citep{LGS11}. In this case, we only need the highest ranked experts as these are most relevant to the query.
\item When a new document reaches the system for the first time (a situation modeled with a long query), the aim is to decide which experts should receive the document. This is a filtering problem \citep{HSS01} and here the aim is to find every relevant person irrespective of their ranking.
\end{itemize}

Although both of these problems might well be regarded as "\emph{the two sides of the same coin}" \citep{BC92} and tackled with a similar approach, in this paper we shall show that differences do exist between them in terms of how they are both formulated and their solutions.

In this paper, we shall consider that the expert's field of expertise is not normally limited to a single subject: a scientist, for example, although specialized in information retrieval, might also have published papers along different research lines (e.g. retrieval models, personalization, recommender systems, etc.) or a politician might sit on three different parliamentary committees (e.g. agriculture, environment and economy) with interventions connected with these areas. If a single profile were built from all of the experts' documents, all of their topics of interest would be mixed up in it. This might result in more general topics taking precedence over more specialized topics and so the profile would not correctly reflect the expert's interests and might mean that they are not found when a specific topic is searched for. One solution might therefore be to consider that a profile is seen not as a monolithic but as a multi-faceted structure comprising other profiles or subprofiles, with each relating to different topics. In this way, the politician would therefore be represented by three subprofiles.

Along these lines, the authors of this paper have undertaken research to find relevant people in a parliamentary setting. In an initial approach, profiles were built for Members of Parliament (MPs) from their parliamentary speeches which could then be used to find relevant MPs \citep{CFH17}. Their profiles were created by considering all of their interventions to build a monolithic profile for each MP. Since many of the MPs' speeches are from specialized parliamentary committees, in \citep{CFH17b} compound profiles were considered whereby each MP could have various subprofiles according to their interventions on the corresponding committees to which she/he belongs. This paper demonstrated that this method of organizing user profiles is much more interesting for the recommendation problem both in terms of profile performance and interpretability.

In this paper, we go a step further because our aim is to determine whether the use of machine learning techniques (and more specifically clustering) could enable the different topics that users are interested in to be automatically discovered and subprofiles to be built on the basis of these. This automatic discovery of topics (groups) would be particularly useful when there is no explicit association of documents or if there is, it is not the best one for optimal performance in recommendation or filtering tasks (topics that should be separated are grouped together in the same subprofile), something which is quite common in a parliamentary context. For example, if we were to consider a parliamentary committee that was created for political reasons to simultaneously cover the three areas of agriculture, livestock and fishery, then all MP interventions on this committee would be included in the same subprofile although they might represent different topics. In addition, the committee structure usually changes with each term of office, and so clustering the MPs' interventions according to these commissions would provide at any given time a topic distribution that depended on organizational political decisions. Finally, the cold start problem at the beginning of a term of office, whereby no committees exist yet, would be reduced by considering the clustered topics learnt from the previous term.

In this paper we show how clustering is a suitable technique for discovering hidden topics from documents and creating compound profiles to represent user interests. Our experimental results also show how clustering techniques may be successfully applied to expert recommendation and filtering problems to build multi-faceted profiles, where each subprofile is obtained from the documents that are relevant to a user and which are grouped together. These two problems can be solved from a unified perspective because conceptually in both contexts, given a query, the result is a ranking of expert users to be recommended or to recommend to. We have also investigated two ways of applying clustering to the set of documents: a global approach, where clustering is carried out by considering all the experts' documents; and a local one, where clustering is only performed with each expert's documents.

In order to describe how clustering is applied to these problems and its performance, this paper is organized as follows: Section \ref{section:Preliminaries} presents introductory information about user profiling and clustering in order to contextualize the rest of the paper; Section \ref{section:profiles} contains the core of the article and describes the clustering proposal for building subprofiles; Section \ref{section:Evaluation} describes the experimental design and the corresponding results and discusses the main findings; Section \ref{section:related} reviews the state of the art, presents similar approaches and examines the differences between these and our proposal by highlighting our contributions; and finally, the last section outlines our main conclusions and future lines of research.

\section{Preliminaries} \label{section:Preliminaries}

Given that the context of this paper is to combine the construction and use of profiles for information access and the application of clustering methods to more accurately organize such profiles, in this section we shall present some concepts and techniques related to these two topics and their combination. Section \ref{section:related} will present a detailed review of the state of the art.

\subsection{User Profiling} \label{subsec:Profiling}

A profile could be defined as a representation of a user model, storing the user's basic information (e.g. age, gender or location), knowledge, background and skills, behaviour and interaction, contextual information, interests or preferences and intentions \citep{SA09,GLW10}. The process of learning a profile is known as user profiling and is based on collecting information explicitly (users express their interests or preferences unequivocally \citep{GSCM07}) or implicitly (a system is in charge of automatically detecting the information items of interest to the user by basically analyzing browsing data).

This paper focuses on profiles that mainly express interests so an adequate method is needed to represent them both efficiently and effectively. Gauch et al. in \citep{GSCM07} consider that profiles could generally be represented by keywords, semantic networks or concepts. Intelligent techniques based on machine learning and data mining, meanwhile, are also applied to represent user models \citep{SA09}. Focusing on keyword-based profiles, they store a list of relevant words extracted from the sources used to build them (documents, web pages, textual descriptions of any type of items, etc). These keywords or terms are weighted in order to reflect their importance for the user and usually modeled as weighted vectors (e.g. by using a TF-IDF weighting scheme \citep{LGS11}). Interests may also be expressed as abstract concepts rather than keywords. More elaborate profile representations that are built by combining different elements (e.g. topics and keywords) will be discussed in Section \ref{section:related}. Although knowledge-based profiles can be obtained (possibly as a human readable representation of user interests), they are not successful for recommendation or filtering problems particularly when it comes to documents that represent speeches and oral discussions.

Profiles are considered basic tools for user adaptation in a wide range of fields in computer science \citep{GLW10} and more specifically, \citep{SA09} indicate various domains relating to information access. Taking into account the context of this paper, these include personalized information retrieval \citep{GZOW13}, recommender systems \citep{BOHG13} and expert finding \citep{LHWL17}.

\subsection{Text Clustering} \label{subsec:Clustering}

From a general point of view, the main purpose of cluster analysis is to attempt to find a common structure over the instances of an unlabeled data set in order to split them into groups (clusters) with similar characteristics \citep{KR90}. When the objects to be clustered are texts this process is called Document Clustering.
The first time this machine learning technique was used in IR was more than 40 years ago, with the aim of improving the efficiency of the retrieval process, originating the {\it Cluster-based retrieval model} \citep{JR71}. Once documents are clustered and related documents are placed in the same group, given a query submitted by a user, this is confronted to the representatives of the clusters and the system would return the documents belonging to those clusters whose representatives are the closest to the query. The fundamental assumption to apply this cluster-based retrieval model is the {\it cluster hypothesis}, stated as ''closely associated documents tend to be relevant to the same requests'' \citep{Rij79}. 

\begin{figure}[tb]
\centering{
\includegraphics[width=1.0\textwidth]{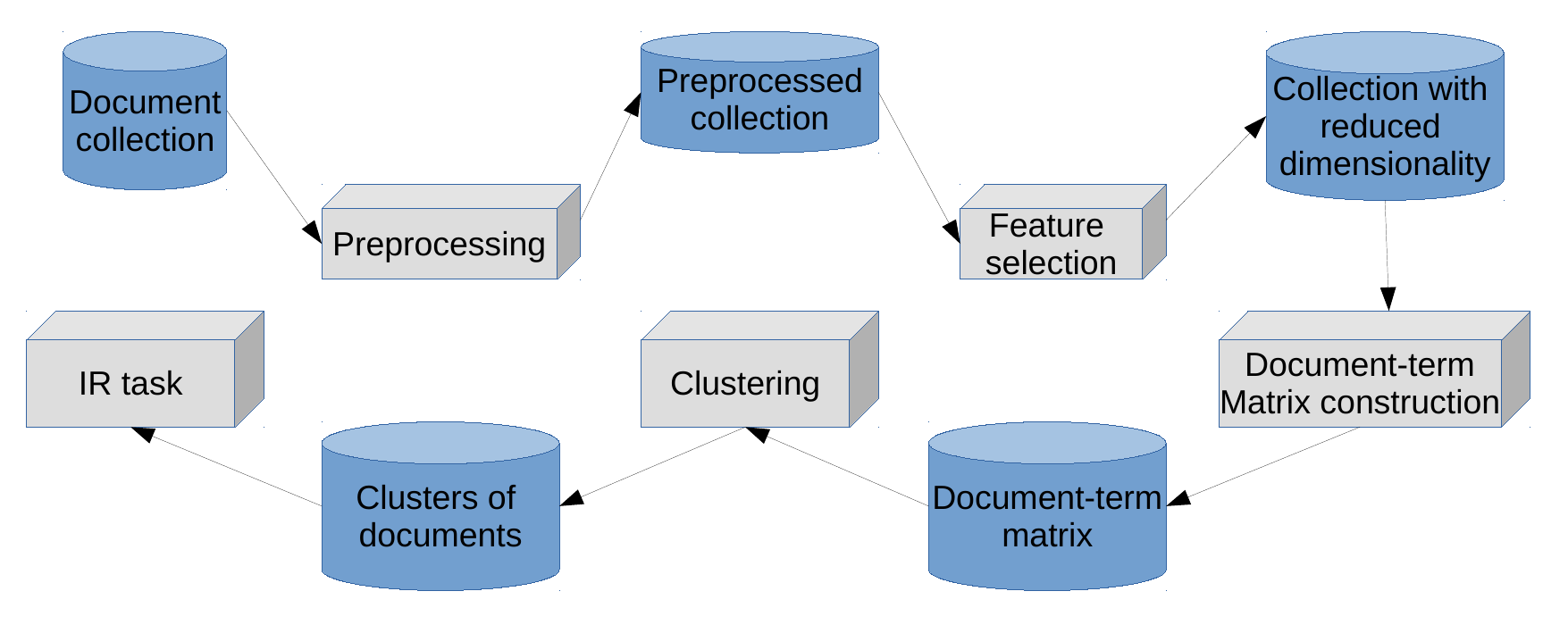}
\caption{Steps in the text clustering process}
\label{fig:procesoClustering}
}
\end{figure}

Figure \ref{fig:procesoClustering} shows the general process of clustering applied to IR. Given a collection of documents, where clustering is going to be performed on, the first step is its preprocessing, which may consist of stop word removal and stemming (removing word suffixes and leaving the words in their lexical stems). The next step could be the reduction of dimensionality of features (the terms), because we are dealing with a high-dimensionality problem \citep{Zamora17}. The construction of the document-term matrix is the next process of the pipeline. The rows corresponds to the documents in the collection and the columns to the terms. If a document contains a term, then in the corresponding cell there will be a weight that reflects the importance of this term in that text. This matrix, which is usually very sparse, will be the input of the clustering algorithm, as well as the number of clusters to generate. As output it will offer a partition of the corpus in such number of clusters. These clusters could be applied in lots of IR tasks \citep{SK11}, for example, document organization and browsing, text summarization, document retrieval, etc.

Of all the various existing clustering techniques \citep{RM05,SM12}, we should highlight two main families. The first of these is connectivity-based clustering or hierarchical clustering \citep{KR90,Roux18,JJ11,ZK05}. This builds a distance tree (or dendogram) to represent the fact that items in the same branch are more similar than items in other branches according to how close they are. This first family is divided into two different categories according to how the dendogram is built: the agglomerative approach \citep{SMA13}, where each instance belongs to an independent cluster at the beginning and pairs of similar clusters are combined recursively in the same way as the agglomerative nesting algorithm (AGNES,  \citep{KR90}), and the divisive approach \citep{Jayaprada14}, where all the instances start in a unique cluster which is separated recursively into two different groups according to similarity as in divisive analysis clustering (e.g. DIANA \citep{KR90}).

The second family is centroid-based clustering. In it, the different clusters are shaped around a middle point which is not necessarily an instance of the data set and each item is assigned to the cluster whose middle point is nearby \citep{macqueen1967,RSJ13}.  We shall focus on two different methods to compare the behaviour of the data in different approaches. The K-Means \citep{LS82,YIHU18} algorithm works by splitting $n$ instances into $k$ different groups and assigning each instance to the group with the nearest mean iteratively and recalculating the group mean point after each iteration. The PAM \citep{KR90,Nguyen17} algorithm function is also similar to the previous one although the middle point of the clusters in PAM is an instance which represents the cluster median.

In addition to this set of classic clustering methods, we may find in the literature other techniques that, not being exactly clustering algorithms, try to capture the underlying semantic of the data and can be adapted or applied to this problem. A first example, in the context of text document collections is Latent Dirichlet Allocation (LDA) \citep{BN03,Bui17}, which is an algorithm that is mainly used in natural language processing. LDA is a three-level hierarchical Bayesian model. It finds the latent topics from a document collection and assigns a probability distribution of topics to each document and also a probability distribution of terms to each topic.
Other example is the Self-Organizing Maps (SOM), which are an effective tool that provides a data visualization of high dimensional space by reducing the dimensions of the data to a low (typically two) dimensional map. SOM implements an artificial neural network that is trained with an unsupervised data set with the objective of condensing all the information of the train set, while the most important topological and metric relations among data are preserved, creating some kind of abstraction of the input space \citep{Kohonen2001}.  

In data clustering analysis, an important problem is to establish the number of clusters and how to calculate it. There are many ways to estimate the number of clusters that best fit the data set. It is highly common in well-known problems to naturally determine the number of clusters in order to obtain a number of well-defined groups but in other cases this is very difficult because there are no clues about this number. More specifically in text databases, an alternative approach to determine the number of cluster is to consider the values of $n$ (the total number of documents), $m$ (the total number of terms) and $t$ (the number of non-zero entries in the respective document-term matrix). The number of clusters $k$ is then defined as $k = mn/t$ \citep{CO90}. Finally, another outstanding approach to determine the value of this parameter is to calculate it with the general and effective method $\sqrt{n/2}$ \citep{KR90}.

With respect to the evaluation of the quality of the clustering process, typical evaluation measures try to maximise the intra-cluster similarity, i.e., documents placed in the same cluster must be very similar among them, and minimise the inter-cluster similarity, i.e. documents placed in different clusters must be very dissimilar. This is the case of the well known Silhouette index \citep{Starczewski15}, which computes the average distance of a given object with the objects of the nearest cluster and subtracts the average distance of an object with respect to the elements from its own cluster (averaged for all the objects). Other example is the Davies-Bouldin index \citep{DB79}, which is the ratio between the within cluster distances and the between cluster distances (averaged as well). It identifies how compact and well separated are the clusters. These are known as internal validity measures because they are computed only with the information of the dataset and the resulting clustering. The other alternative is to perform an external evaluation that depends on the application domain. In those cases where clustering is only part of the system being built, it is important to evaluate how the clustering algorithm affects the global behaviour of the system \citep{Croft15}. As this is our case, the clustering quality will be indirectly measured through the quality of the obtained recommendations (using standard measures in this IR field -- see Section \ref{section:Evaluation}).


\section{Building multi-faceted profiles by clustering documents} \label{section:profiles}

As we mentioned in the introduction to this paper, since a user might be interested in a number of different topics and their profile consists of a set of concepts or topics comprising weighted terms, we could in turn say that the profile is multi-faceted since it attempts to capture the different facets contained in the set of documents associated to a user. In this paper, each facet or concept comprising a profile will be called the subprofile. These multi-faceted profiles are the opposite of monolithic profiles where the underlying topics are not made explicit.

In most situations, the concepts are hidden, i.e. they are implicit in the set of documents. This means that a process to automatically extract or learn them is required. In our case, we have applied clustering analysis. The idea is to cluster the sets of documents to obtain $k$ groups of documents. The input of the cluster algorithm is a matrix where the rows are the instances, i.e documents, and the columns are the attributes, the terms from the vocabulary of the document collection, taking real values that usually represent the importance of the term in the document ($0.0$ when a term does not occur in a document). The number of desired clusters is also an input for the algorithm. The output is a set of clusters, and within each one, there is high similarity between all the documents (we could say that all the documents in the clusters deal with the same topic) but low similarity with the documents from other clusters. We assume that each group represents a concept and by combining all the terms from the included documents, its final representation is a list of weighted terms. 

For the purposes of creating user profiles based on the content of their documents, we could consider two approaches for clustering their documents. The first is a local approach and finds the underlying document groups locally for each user, i.e. by only considering their documents. The alternative option is global because it performs the clustering process with all the documents from every user. The first approach captures a specific user's topics whereas the second attempts to find common concepts that are usually shared by every user. This means that in local clustering, the learned groups are exclusive for each user and therefore only contain documents for that user. In global clustering, the clusters will contain documents from different users. A subprofile for each user will therefore be obtained from this global clustering by grouping the documents within each cluster that belong to the given user.

\begin{figure}[tb]
\centering{
\includegraphics[width=1.0\textwidth]{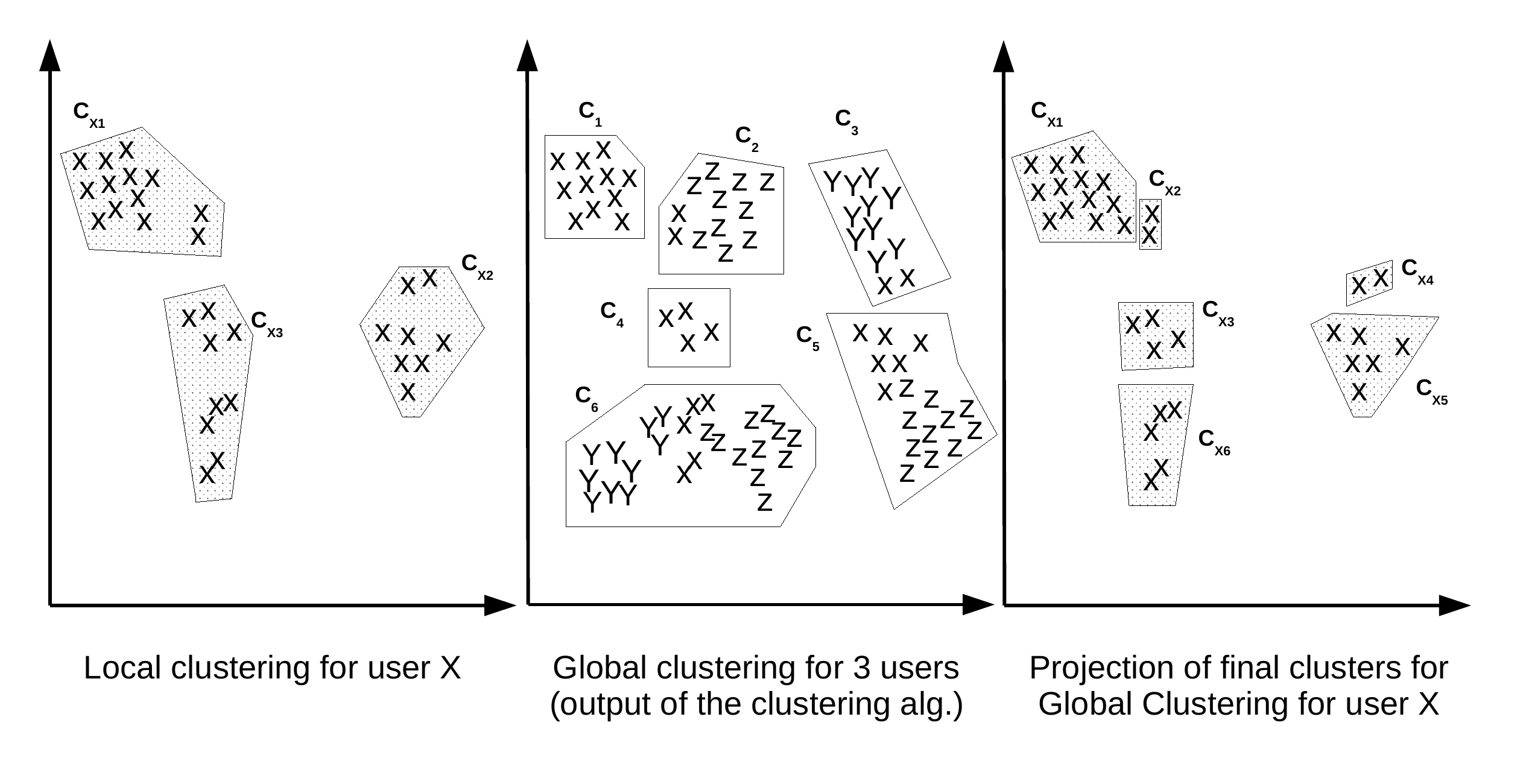}
\caption{Example of local and global clustering}
\label{fig:localGlobal}
}
\end{figure}

The left-hand side of the graph in Figure \ref{fig:localGlobal} shows the arrangement of all of user X's documents and how they are grouped into local clusters of similar documents. From this aggregation, three subprofiles will be built for the user. In terms of the global approach where the documents of all the users (X, Y and Z) are incorporated into the clustering algorithm, the central graph shows the hypothetical groups found. The clusters $c_2$, $c_3$, $c_5$ and $c_6$ are heterogeneous in the sense that they integrate documents from different users. If we again focus on user X, the number of profiles to be built following this global approach will depend on the number of clusters the documents belong to. On the right-hand side of the graph, we can see that new clusters are considered for X and so the final number of clusters for X is $6$, and this will therefore be the associated number of subprofiles for this user.

In both cases, the output of the non-supervised process is an association of each document from the given user to a cluster. For each user and for a given cluster, a "macro-document" is created by compiling all the documents included in that same cluster. This document will correspond to a subprofile. A new document collection is generated containing all the subprofile documents from all the users. This will be indexed for use by the corresponding information retrieval system. When a query is submitted to the system, it returns a ranking where different subprofiles from the same user may be distributed across it. As we are recommending experts, the final ranking must be composed of users, so it becomes necessary to use some fusion strategy to compute a final score for each user, considering all their different subprofiles in the ranking.

\section{Evaluation} \label{section:Evaluation}

This paper addresses the general problem of finding people but our evaluation will focus on a parliamentary setting. The basic objective is to find relevant Members of Parliament (MPs) given a query formulated by a citizen or to determine which MPs might be interested in reading a new document received by the system. In order to do so, we have opted to represent MP interests by means of a profile that will be constructed on the basis of their interventions in the political initiatives presented at the public parliamentary sessions. More specifically, let us take into account the fact that an MP might sit on a number of committees and that these are smaller in terms of the number of MPs involved and cover more specific topics. Since the user might be interested in several political topics (e.g. agriculture, education, economy, etc.), the aim is to create subprofiles for each MP to represent the MP's interest in these different topics.

The general objective of this evaluation is to determine whether text clustering is a good tool for automatically identifying the different topics of interest to a user and whether it might be useful to recommend experts to them and filter information for them. In order to achieve this, we propose that the following specific research questions be answered by means of the evaluation described in this section:

\begin{itemize}
\item RQ1: Is text clustering an appropriate technique to automatically extract the topics in which a person is interested by considering the particular features of the parliamentary context?
\item RQ2: Do filtering and recommendation tasks benefit from clustering-based subprofiles?
\item RQ3: Is there any difference between building the clusters locally and globally?
\item RQ4: Is the number of clusters relevant for recommendation quality?
\item RQ5: What are the best clustering algorithms for these tasks?
\end{itemize}

In this section, we shall therefore describe the experimental design and also the results of the experiments that have been conducted in this evaluation stage.

\subsection{Test Collection}

The dataset that we have used for the experiments is the collection of Records of Parliamentary Proceedings from the Andalusian Parliament in Spain and more specifically those that belong to 8th Term of Office\footnote{Available from \url{http://irutai2.ugr.es/ColeccionPA/legislatura8.tgz}}. This has been organized around the initiatives discussed in committee and plenary sessions containing a total of 5258 records with 12633 interventions. There are 26 different committees and a total of 132 spokespersons. For experimental purposes, we have selected only those MPs with at least 10 interventions.

\subsection{Overview of the Recommender and Filtering System} \label{section:system}

In order to recommend MPs given a citizen's query or a document to be filtered, we have used the open source Apache Lucene Library\footnote{https://lucene.apache.org/}, implementing the well-known BM25 model as a retrieval model \citep{JWR00}. For each $MP_i$, the input of the indexer is the set of their subprofiles. For example, the documents to be indexed for $MP_5$ are the subprofiles for three clusters, $c_1$, $c_2$ and $c_3$ (therefore 3 in total, called $MP_5\_c_1$, $MP_5\_c_2$ and $MP_5\_c_3$, respectively). The terms contained in these are filtered, the stop words removed, and reduced to their roots using the stemmer implemented in the Lucene Spanish Analyzer. Any term occurring in fewer than $1\%$ of the interventions is then removed. Given a query, a ranking of MP subprofiles is given as output. However, as the final objective is to rank MPs according to their relevance to the query, the original ranking is filtered by considering the $CombLgDCS$ method presented in \citep{CFH17b}. This strategy calculates a single score for each $MP_i$ by aggregating the different scores of their subprofiles but logarithmically devalued according to their positions in the ranking. The formula is the following:

\begin{equation}
 score(MP_i,q) = \sum_{MP_i\_c_j} \frac{s(MP_i\_c_j)}{\log_2(rank(MP_i\_c_j) +1)}, 
 \end{equation}

\noindent where $MP_i$ is an MP, $MP_i\_c_j$ is a subprofile in the ranking of this politician, $s(MP_i\_c_j)$ denotes its score value (similarity between the profile and the query $q$) and $rank(MP_i\_c_j)$ is the position of the $MP_i\_c_j$ subprofile in the ranking.

Once the scores have been computed for every MP, they are ranked accordingly.

\subsection{Clustering Algorithms}

In our experiments, we have tested the R implementations of the following clustering algorithms: AGNES and DIANA as hierarchical methods (agglomerative and divisive, respectively), K-Means and PAM as centroid-based methods, and finally latent Dirichlet allocation (LDA) and Self Organizing Maps (SOM) as generative statistical model-based and artificial neural networks-based methods, respectively. These algorithms have been selected due to the fact that they are state-of-the-art clustering methods or have been used in the clustering process.

These algorithms obtain a matrix as the input with documents for the rows and terms for the columns. After a process of stop words, number removal and stemming, the remaining terms are weighted using the TfIdf scheme. The documents are therefore represented by vectors containing the different terms in the collection for the columns. If a term occurs in the document, then the weight in the corresponding position is the TfIdf, and $0.0$ otherwise.

In the local approach for a given MP, $MP_i$, the number of instances is equal to the number of documents associated to her/him. The clustering process is repeated for each user in the system so all the users will obtain their own clusters. In the global method, on the other hand, the number of instances is the number of documents in the system and the clustering algorithm is executed just once, obtaining a group of clusters for every MP as the output.

Both centroid-based and hierarchical methods use cosine dissimilarity to compute the distance between individuals. In terms of the LDA algorithm \citep{BN03} used for clustering, once the algorithm has found the distribution of topics for all the documents, each document is assigned to the cluster associated to its most probable topic. With respect to SOM, it can also be used in order to group similar data together. Once the SOM output is obtained, and each document is associated to a neuron, there is a set of weights vectors which represent the position of the neurons in the discretized space of the data and those vectors are grouped in function of their similarity using any clustering method, thus creating clusters of similar instances of the real data that are attached to the clustered neurons. In our case we have used SOM in combination with the K-Means algorithm (noted as SOM-KM) as it has been found as a state-of-the-art association in general clustering tasks \citep{PPP11,PGB16,JLLH13}.

\subsection{Selecting the Number of Clusters}

As we have already mentioned, the number of clusters, $k$, given as the output is an important issue in any problem where clustering is applied. The ideal situation is the automatic selection of the best possible value but this is not easy.

In our experimentation, we have tried different approaches, where $k$ is fixed or is computed automatically by taking into account some collection-dependent data. More specifically, we have conducted experiments with the following alternatives:

\begin{itemize}
\item $k=\#Com \Rightarrow$ For global clustering, this represents the number of committees in the eighth Term of Office of the Andalusian Parliament, i.e. $26$. For local clustering, this number is specific for each MP, and is the number of committees in which each MP has participated: $6.02$ committees on average with a standard deviation of $4.52$. The objective of setting this value to $k$ is to determine the degree to which the clustering algorithms are able to reproduce the groups of parliamentary initiatives given by the official committees, which is considered as the ground truth.

\item $k=m*n/t \Rightarrow$ $m=$ number of terms in the Andalusian Parliament collection; $n=$ number of interventions in the collection; and $t=$ number of non-zero entries in the document-term matrix. This is applicable to both clustering approaches, although the values of $m$, $n$ and $t$ will depend on the corresponding type. In the case of global clustering, $m$ is $4208$; the total number of MP interventions ($n$) is $10025$ ($80\%$ of the total number of interventions (the training partition) and $t=1,702,296$. For local clustering, these numbers vary because they depend on the number of each MP's interventions, but on average, $m=3427.45 \pm 2056.15$, $n=58.11 \pm 58.55$ and $t= 12106.66 \pm 12064.64$. The final value for $k$ for the global approach is $k=24$, and for the local approach, the average is $15.85 \pm 9.67$.

\item $k = \sqrt{n/2} \Rightarrow$ For global clustering, this value is $70$, computed by considering $n=10025$ ($80\%$ of the total number of interventions --the training partition), while for the local one, as the number of interventions of a given MP is specific for each politician, the mean value is $4.25 \pm 2.60$.

\end{itemize}

\subsection{Experimental context} \label{subsec:experimentalContext}

The set of initiatives is randomly partitioned into a training set (80\%) and a test set (20\%). The training set is used to build MP subprofiles starting with the clusters obtained and the test set is used for evaluation purposes. This process is repeated five times, and in this paper the reported results are the average values.

We shall use the content of the initiative (full text) as the query for the filtering process (in this case, our aim is to distribute an initiative to any MP that might be interested), and the initiative title for the case of the MP recommendation approach (the aim is to find an MP to talk to, for example, so we might want to obtain the highest ranked relevant MPs). In both cases, and focusing on relevant judgments, since the objective is to find MPs who might be familiar with the topic, the ground truth for each query will only comprise those MPs who participate in its corresponding initiative. Since it is quite reasonable to assume that an initiative will also be relevant and of interest to other MPs who may not have participated in it, we could say that this is a rather conservative assumption to evaluate, particularly for the filtering task.

Given a query, the search engine will return an MP ranking. In order, therefore, to measure the quality, we will use the well-known precision and recall metrics, focusing on the top 10 results (p@10 and r@10, respectively). We will also consider normalized discounted cumulative gain \citep{JK02} (ndcg@10) in order to consider the ranking position of the relevant documents.

In order to ascertain whether learning the subprofiles is a good approach for representing the MP profile, we have opted to compare the results with three different baselines:

\begin{itemize}
\item A single profile for each MP (monolithic profile). From all of the MPs' interventions on all of their different initiatives, only one profile is built for them. This profile will contain all the topics in which they are interested. We could say that this is the case where $k=1$.
\item Several subprofiles are built for each MP according to the committees they are involved with (committee-based subprofiles). Each MP will have different associated subprofiles by considering their difference committee interventions. The committee interventions will be the input for building the corresponding subprofile. From a practical point of view, if a given MP has participated on $k$ committees, their profile will comprise $k$ subprofiles.
\item One subprofile for every initiative in which an MP has participated (intervention-based subprofiles). This is the extreme case where each MP's interventions on an initiative will comprise its own subprofile. The number of subprofiles associated with an MP will therefore be the same as the number of her/his initiative interventions.
\end{itemize}

The underlying idea behind these baselines is to have two extreme situations (i.e. one profile for each MP or as many as the number of their interventions) and an intermediate one, where the number of subprofiles is established by the committees on which they participate. The expected situation would be that the MP recommendation and filtering tasks would perform better with clustering-based subprofiles than those obtained by the baselines.

\subsection{Results}

In the following sections, we shall present the results of our experiments and answer the following research questions.

\subsubsection{RQ1: Is text clustering a suitable technique to automatically extract topics in a parliamentary context?}

In order to answer this first research question, we shall show how clusters cover the political topics discussed in the sessions considering both, a qualitative analysis focused  on a particular MP and  a broad qualitative analysis, focusing on committees.

\paragraph{Individual Qualitative Analysis}

This analysis considers an MP from the {\texttt{Izquierda Unida}} party. We selected him because he is a prolific MP (during the 8th Term he  spoke in 172 different sessions) covering a wide range of topics  (besides 97 interventions in plenary sessions  he also participated  in 14  specialized committees or working groups, his remaining 75 interventions). So, which  are the topics that the MP is 'truly' interest in? We can say that they are related to the committees he participated, but it is common that some topics have more strength than others in the  MP's interests. To quantify this idea we can see  the second column in Table \ref{tab:profiles}, where we show the size (in terms of percentage of terms) of his different interventions (note that half of the weight is located into Plenary Sessions, where several topics might be discussed). Note that   from these data we can see that he is focused on  Equality and Social Welfare, Culture  and Health (representing the 70\% of his interventions in committees, i.e. without considering Plenary Sessions).

\begin{figure}[tb]
\centering{
\includegraphics[width=0.45\textwidth]{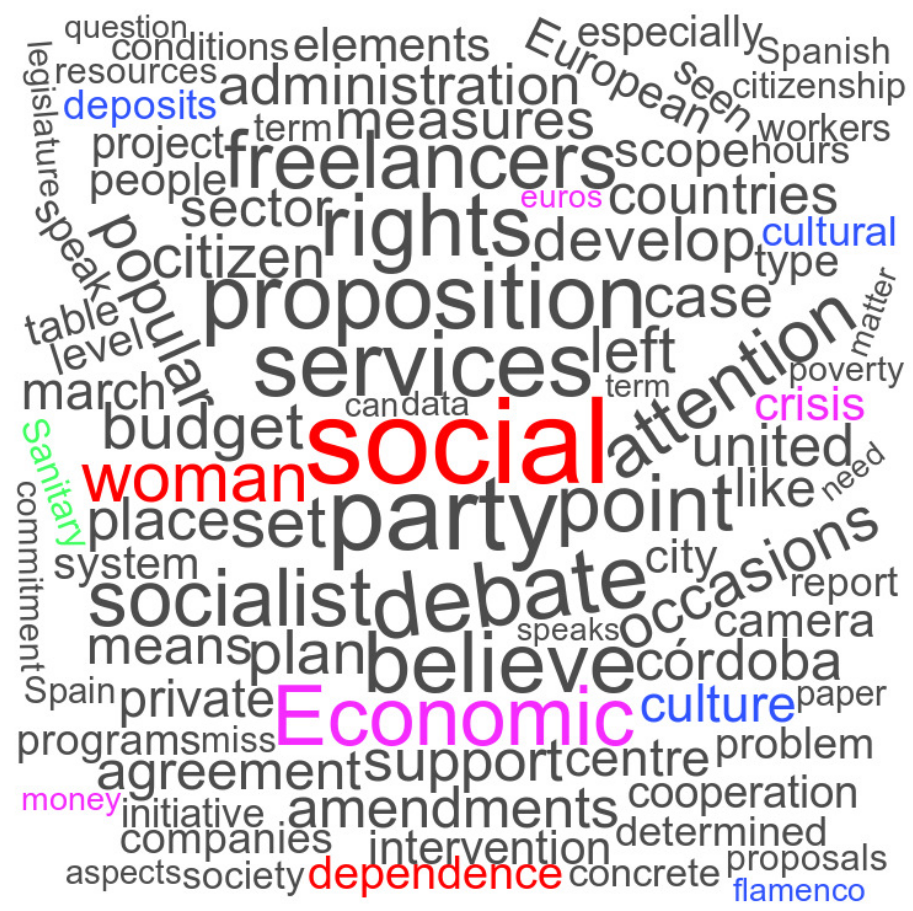}
\includegraphics[width=0.45\textwidth]{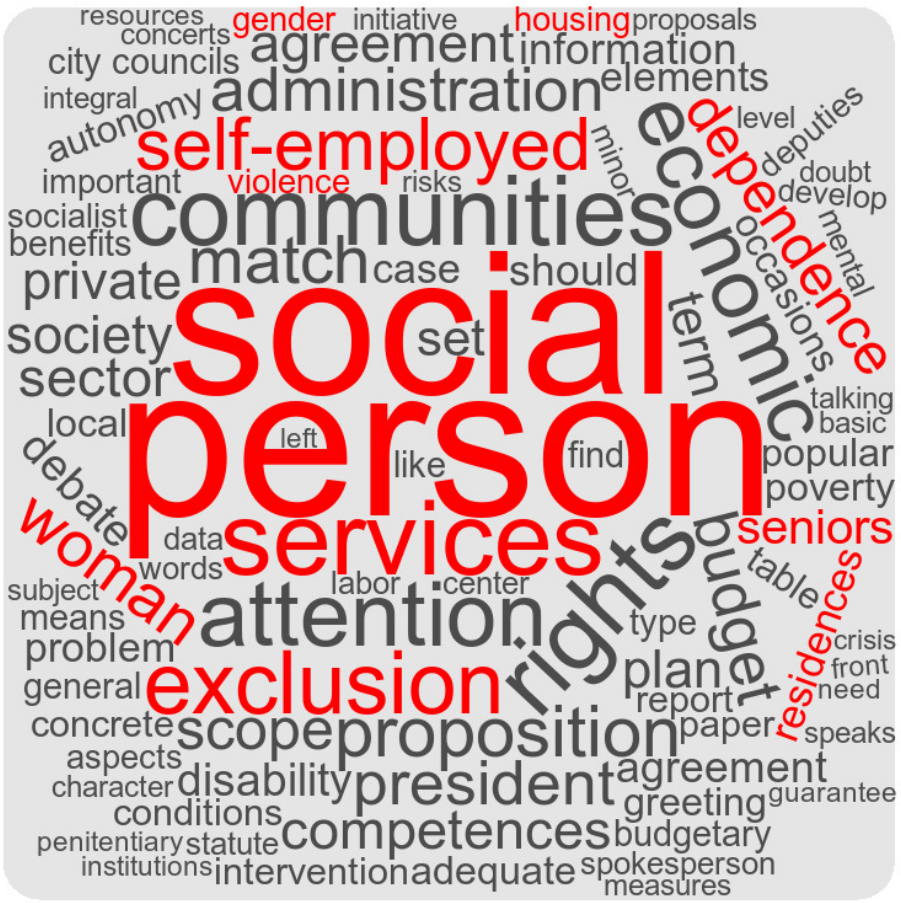}
\caption{Word Cloud representation for the different profiles: left graph shows Monolithic (all the interventions forms a unique profile) and right graph shows a Committee-based profile obtained using the data from the ''Gender Equality and Social Welfare Committee.
\label{fig:profiles}
}}
\end{figure}

Let us consider firstly those situations in which we do not perform any clustering algorithm, i.e.,  monolithic and committee based-profiles.  Focusing on  the monolithic profile, we found that it is dominated by terms related to the parliamentary procedures  being difficult to identify the topics the MP is interested in, as the word cloud on the left hand side in Figure \ref{fig:profiles} shows. On the other hand, if we consider committee-based subprofiles, see for example   the right  word cloud in Figure \ref{fig:profiles}  obtained from the ''Gender Equality and Social Welfare Committee, those terms related to the committee dominate the cluster, although common terms in the parliament appear, but with less frequency.  Focusing on the large number of interventions in plenary sessions, we  do not have any previous association to a given topic, and therefore  they are joined  in a big  profile, exhibiting the same pattern than monolithic-based profiles.

\begin{table}[htbp]
\caption{Distribution (in terms of profile's size) of the MP interventions in the term of office. The second column shows the 'true' distribution considering the real sessions in the parliament. The third column shows the distribution considering the learned clusters.}
\begin{center}
\begin{tabular}{|l|r| r|}
\hline
\multicolumn{2}{|c|}{Real Distribution} & Clustering \\
\hline
Plenary Sessions & 0,500    &  \\ \hline
\hline
\multicolumn{2}{|c|}{Committees} &  \\ \hline
Gender Equality and Social Welfare & 0,128  & 0,286 \\ \hline
Culture   & 0,121    & 0,151 \\ \hline
Health   & 0,103    & 0,144  \\ \hline
Presidency   & 0,046    & \multicolumn{1}{l|}{} \\ \hline
Tourism and Business   & 0,018    & 0,021 \\ \hline
European Affairs   & 0,015  & 0,052 \\ \hline
Public Work and Housing   & 0,011    & 0,013  \\ \hline
Public Work  and Transports & 0,010   & 0,030 \\ \hline
Technology, Science and Business   & 0,009     & 0,063 \\ \hline
Trade, Technology and Science & 0,009   & \multicolumn{1}{l|}{} \\ \hline
Governance   & 0,008    & \multicolumn{1}{l|}{} \\ \hline
Justice  & 0,008    & \multicolumn{1}{l|}{} \\ \hline
Radio and Television   & 0,007     & 0,016 \\ \hline
Environment & 0,005&   \multicolumn{1}{l|}{} \\ \hline
\hline

Economy topic& \multicolumn{1}{l|}{} &  0,139  \\ \hline

Gender violence topic & \multicolumn{1}{l|}{} &   0,066 \\ \hline
 Labour movement topic & \multicolumn{1}{l|}{} &  0,007\\ \hline
 Education topic & \multicolumn{1}{l|}{} &  0,007\\ \hline
 Young people  topic & \multicolumn{1}{l|}{} &  0,006\\ \hline
\end{tabular}
\end{center}
  \label{tab:profiles}
\end{table}

\begin{figure}[tb]
\centering{
\includegraphics[width=0.46\textwidth]{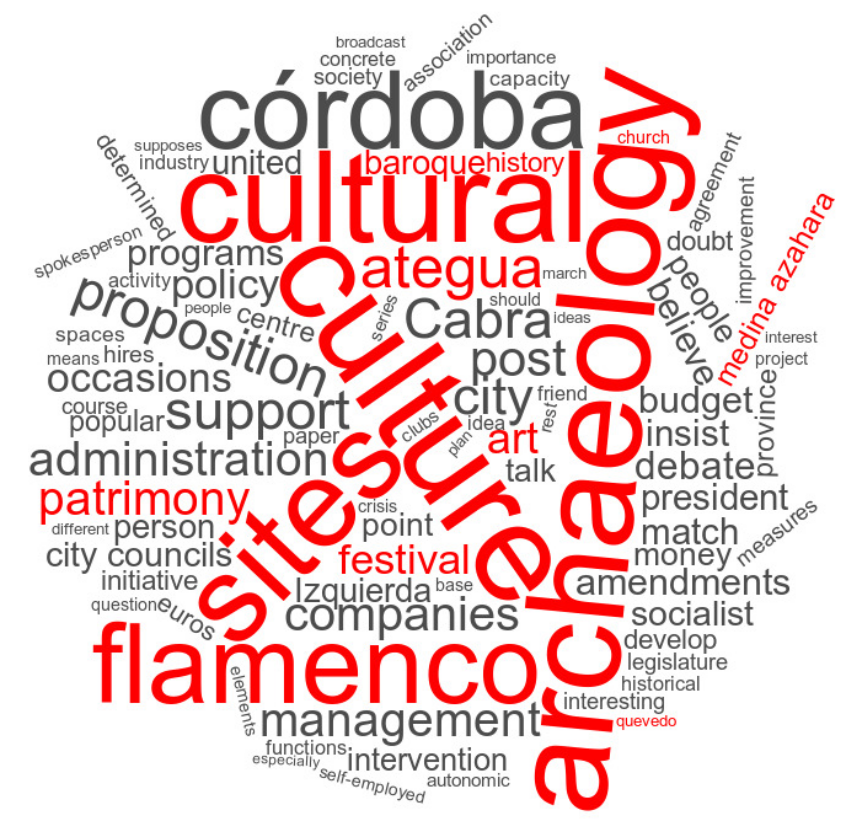}
\includegraphics[width=0.53\textwidth]{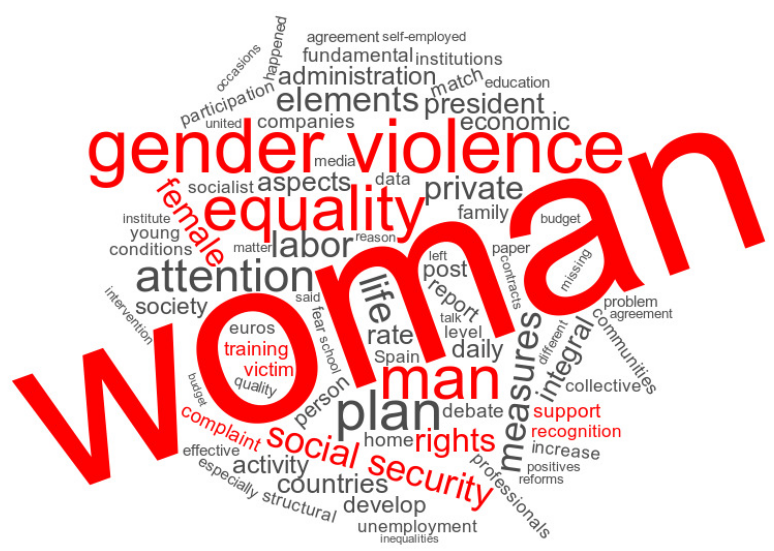}
\caption{Word Cloud representation for two  different learned profiles using Global K-Means as clustering approach.}
\label{fig:profiles2}
}
\end{figure}

Now, we will focus on the results obtained after applying a clustering algorithm, particularly Global K-Means, being the value of $K$ equal  to $26$. In this case, all the interventions of the MP (including plenary sessions) are distributed among 14 of the 26 candidate clusters. The size of each cluster (in terms of percentage of terms)   is shown in the last column of Table \ref{tab:profiles}. In order to identify the dominant topic of each cluster,  a logical approach is to see the most common terms in the cluster, those which  have the highest contribution to it, and   assign the cluster with the topic they suggest, appearing different situations:

\begin{itemize}
\item It is possible to find a 1-to-1 match between the documents in the cluster and a given committee, as is illustrated in the left hand side of  Figure \ref{fig:profiles2} with red words suggesting   that  the  cluster is related to \texttt{culture}.
\item Also, a committee can be split into different topics, 1-to-n. For example, clustering was able to discover {\em ''gender violence''} as a new topic, as the graph in the right-hand side of Figure \ref{fig:profiles2} shows. The interventions in this cluster are highly related to ''Gender Equality and Social Welfare'' committee, but clustering was able to distinguish among  {\em ''gender violence''} and {\em ''social welfare''}.

\item Joining  two different committees in one cluster, 2-to-1: The interventions of two highly related committees as ''Technology, Science and Business'' and ''Trade, Technology and Science''\footnote{The reasons for the existence of different committees with highly overlapping topics are political. These committees do not overlap in time but any governmental restructuring also causes modifications to the committees associated with certain areas},  are grouped in the same cluster. 

\item Discover transversal topics, n-to-1:  there exist  clusters having interventions from several committees, as is the case of the topic of {\em ''economy''}, representing a transversal  interest for the MP. This topic  includes  interventions from plenary sessions and  a large number of committees. This reflects that \texttt{economy} is a multidisciplinary topic  shared by all the political activities,  although it has not been stated explicitly.
\item In other  cases,  a global cluster includes   only one intervention of this MP, so it can be considered they represent a marginal topic for the MP interest (last three rows in Table \ref{tab:profiles}).
\end{itemize}

Therefore, we can say that clustering was able to identify the topics of interest of the MP, beyond the committees he is a member. Moreover, it can help to distribute his  interventions in plenary sessions among the respective topics.

\paragraph{Overall Qualitative Analysis}

\begin{figure}[tb]
\centering{
\includegraphics[width=1.0\textwidth]{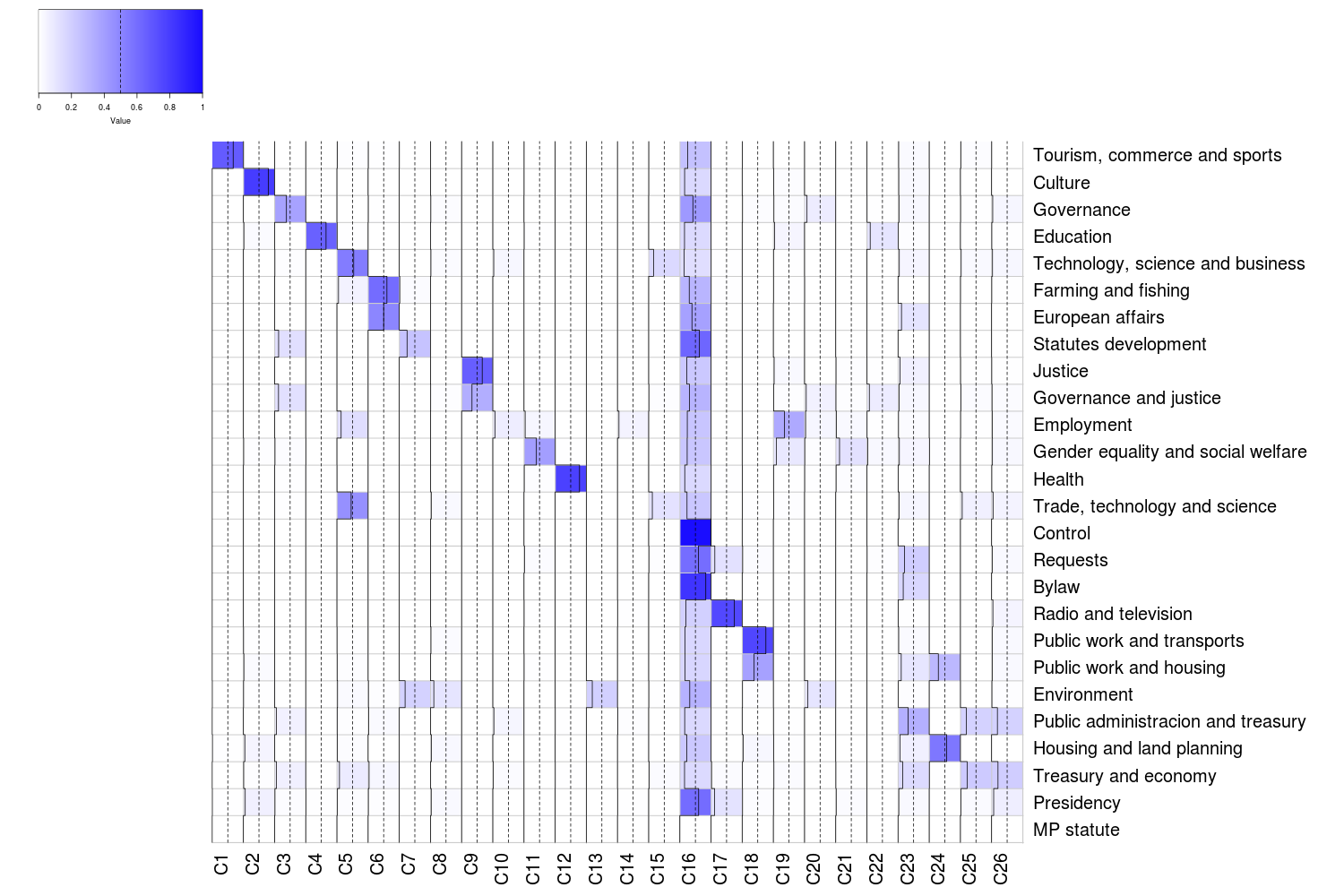}
\caption{Distribution of committees in clusters given by the Global K-Means algorithm}
\label{fig:correspondenciaKMEANS}
}
\end{figure}

\begin{figure}[tb]
\centering{
\includegraphics[width=1.0\textwidth]{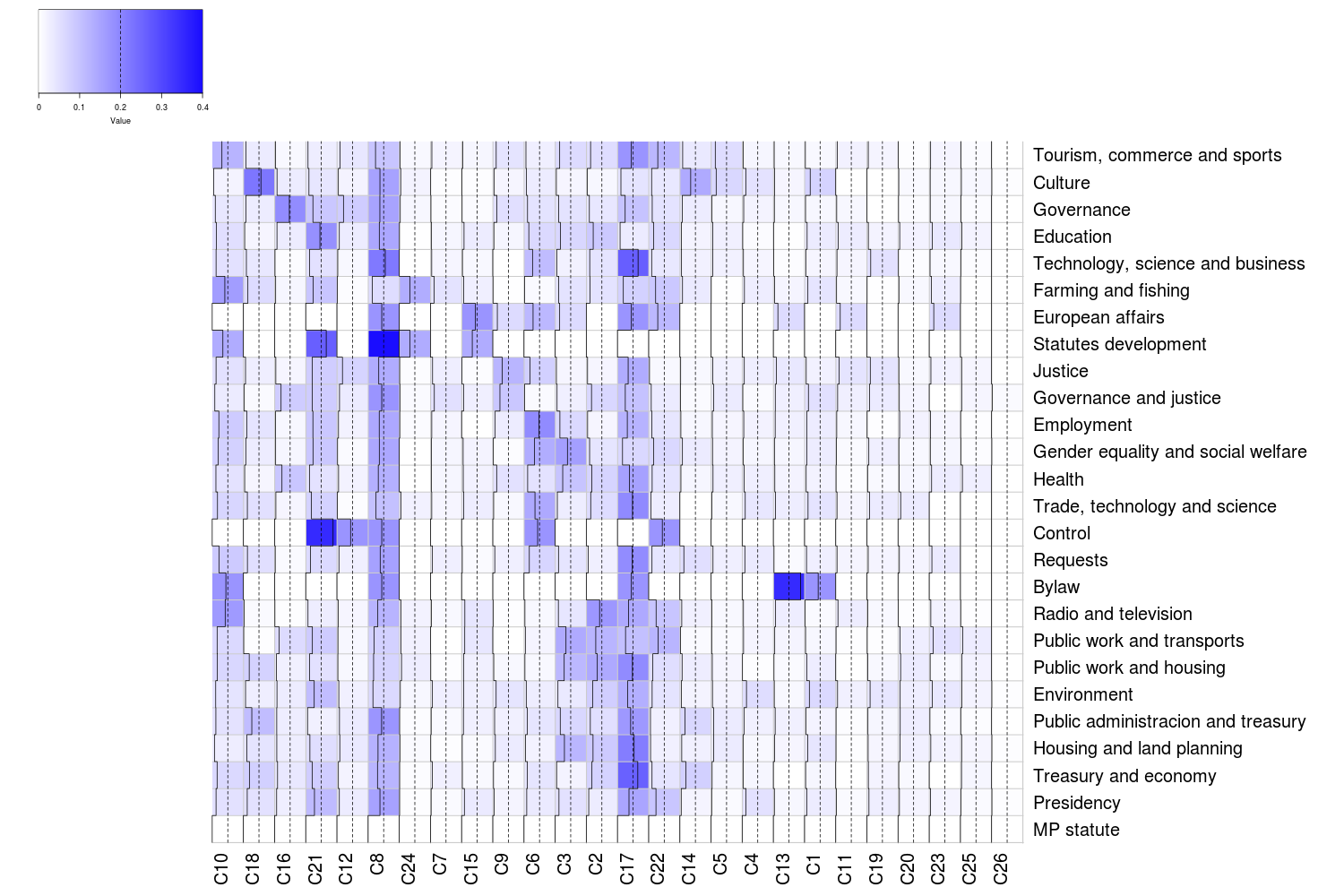}
\caption{Distribution of committees in clusters given by the Global Diana algorithm}
\label{fig:correspondenciaDIANA}
}
\end{figure}

In this analysis the objective is to graphically determine the degree to which each output of the clustering algorithms is close to or far from the ground truth, in our case, given by the official Parliamentary committees which represent a political division of the subjects discussed by the committees. Note that in this case we include all the MPs and all the committees, so clustering were therefore executed with $k=\#Com$, i.e. $26$. In the coloured matrices represented in Figures \ref{fig:correspondenciaKMEANS} and \ref{fig:correspondenciaDIANA}, we have arranged the 26 clusters into columns, and the committees into rows. The shade of each cell provides information about the percentages of the committees included in each cluster (the higher, the darker). We have selected K-Means and DIANA to represent the behaviour of the other algorithms.

Although Figure \ref{fig:correspondenciaKMEANS} represents the Global K-Means, a very similar pattern is displayed by the Global LDA algorithm and we can observe that there is a considerable degree of matching between the different committees and the clusters, i.e. most of the initiatives from the same committee have been assigned to the same single cluster. As mentioned before, in many cases, the clusters have been able to capture the essence of the topics associated to a given committee. This is the case, for example, of Clusters 12 (almost entirely comprising documents from the Health Committee), 2 (associated with the Culture Committee) or 4 (the Education Committee). Also, clustering is able to group together the documents belonging to different committees, which essentially deal with the same topics. This is the case, for example, of the Committee for Public Works and Transport and the Committee for Public Works and Housing, which are mostly included in Cluster 18. In other cases, the same committee is divided into different clusters (for example,  see Clusters 23, 25  and 26) related to economy and public administration.
  Therefore, Global K-Means and LDA capture the subjects from committees with relatively high precision.

Another detected pattern is the one presented by Global DIANA. Figure \ref{fig:correspondenciaDIANA} shows a very different behaviour: considering the committee arrangement given by the K-Means clustering, in order to establish a common comparison point, the diagonal presented in Figure \ref{fig:correspondenciaKMEANS} does not occur in the same noticeable way and the distribution of committees in a single cluster does not occur, so initiatives from the same committee are split into different clusters. For example, documents from the Committee for health are distributed in 10 clusters, as happens with many other committees.

We could conclude that the clustering of official Parliamentary documents better reflects the different topics present in the initiatives than the sometimes artificial political division, and this is probably why the profiles resulting from these clustering processes behave better than those originating directly  from the committees, as shown in Tables \ref{tab:ResultadosFiltrado} and \ref{tab:ResultadosRecomendacion}.

\subsubsection{RQ2: Do filtering and recommendation tasks benefit from clustering-based subprofiles?}

In order to answer this question, and once we have performed the evaluation described in Section \ref{subsec:experimentalContext}, we present the results in Tables \ref{tab:ResultadosFiltrado} and \ref{tab:ResultadosRecomendacion}. The first contains the results of the filtering task and the second shows the results of the recommendation task, respectively. In both tables, the first column indicates the type of clustering ({\it T} (global or local), the second ({\it Alg.}) contains the name of the clustering method, and the third the method for computing $k$, and its value ({\it \#Clusters}\footnote{For local clustering, as this \#Clusters depends on each MP, we show the mean and standard deviation of every MP}). Below the four columns, we have also included the baselines: monolithic profiles (M-Prof) and committee and intervention-based subprofiles (C-SubP and I-SubP), respectively. The columns labeled with {\it r@10}, {\it p@10} and {\it ndcg@10} contain the values of the Precision, Recall and NDCG metrics for the top 10 documents. As considering different measures would give different rankings of methods and baselines, we have attempted to find a way to show a final ranking that would unify these three metrics with a clear idea of the overall performance of the compared methods. We have therefore used Reciprocal Rank Fusion as presented in \citep{CCB09} and which is originally a method for combining rankings from different IR systems to offer a single ranking. We have therefore computed the position of each clustering method and baselines in the ranking resulting from each metric (in the tables, columns {\it P-r}, {\it P-p} and {\it P-ndcg}, respectively) and the value of the Reciprocal Rank Fusion ({\it RRF} in the tables). Finally, we ranked the methods in decreasing order according to this last value and presented the clustering approaches and the baselines according to this new ranking. We believe this fairer way of presenting the results facilitates analysis and enables conclusions to be drawn.

\begin{table}[htbp]
\caption{Values of the evaluation metrics for profiles based on clusters and baselines for filtering (Labels of columns: T = Type of clustering; (L)ocal or (G)lobal; k = method for computing the number of clusters; $\#Clusters$ = number of clusters; r@10 = recall at 10; P-r = Position in the recall ranking; p@10 = precision at 10; P-p = Position in the precision ranking; ndcg@10 = NDCG at 10; P-ndcg = Position in the NDCG ranking; RRF = Reciprocal Rank Fusion value)}
\begin{small}
\begin{tabular}{|l|l|l|l|c|c|c|c|c|c|c|}
\hline
\multicolumn{1}{|c|}{\textbf{T}} & \multicolumn{1}{c|}{\textbf{Alg.}} & \multicolumn{1}{c|}{\textbf{k}} & \multicolumn{1}{c|}{\textbf{\#Clusters}} & \multicolumn{1}{c|}{\textbf{r@10}} & \multicolumn{1}{c|}{\textbf{P-r}} & \multicolumn{1}{c|}{\textbf{p@10}} & \multicolumn{1}{c|}{\textbf{P-p}} & \multicolumn{1}{c|}{\textbf{ndcg@10}} & \multicolumn{1}{c|}{\textbf{P-ndcg}} & \multicolumn{1}{c|}{\textbf{RRF}}  \\ \hline
G & AGNES & $\sqrt{n/2}$ &  $70$  & 0.7724 & 1 & 0.1779 & 1 & 0.6549 & 2 & 0.0489 \\ \hline
G & AGNES & $\#Com$ &  $26$  & 0.7660 & 4 & 0.1754 & 7 & 0.6547 & 3 & 0.0464 \\ \hline
G & PAM & $\sqrt{n/2}$ &  $70$  & 0.7698 & 3 & 0.1767 & 3 & 0.6391 & 10 & 0.0460 \\ \hline
G & AGNES & $m*n/t$ &  $24$  & 0.7652 & 5 & 0.1752 & 8 & 0.6543 & 4 & 0.0457 \\ \hline
L & DIANA & $\sqrt{n/2}$ &  $4.25 \pm 2.60$  & 0.7630 & 6 & 0.1766 & 4 & 0.6379 & 12 & 0.0447 \\ \hline
L & AGNES & $\sqrt{n/2}$ &  $4.25 \pm 2.60$  & 0.7710 & 2 & 0.1775 & 2 & 0.6308 & 22 & 0.0445 \\ \hline
G & DIANA & $\#Com$ &  $26$  & 0.7567 & 12 & 0.1744 & 13 & 0.6509 & 5 & 0.0430 \\ \hline
L & KMEANS & $\sqrt{n/2}$ &  $4.25 \pm 2.60$  & 0.7567 & 13 & 0.1749 & 9 & 0.6408 & 8 & 0.0429 \\ \hline
L & SOM-KM & $\sqrt{n/2}$ &  $4.25 \pm 2.60$  & 0.7559 & 14 & 0.1754 & 6 & 0.6377 & 14 & 0.0422 \\ \hline
G & PAM & $m*n/t$ &  $24$  & 0.7613 & 7 & 0.1748 & 11 & 0.6366 & 16 & 0.0422 \\ \hline
L & LDA & $\sqrt{n/2}$ &  $4.25 \pm 2.60$  & 0.7595 & 10 & 0.1761 & 5 & 0.6303 & 23 & 0.0417 \\ \hline
G & DIANA & $m*n/t$ &  $24$  & 0.7545 & 16 & 0.1739 & 16 & 0.6470 & 6 & 0.0415 \\ \hline
G & PAM & $\#Com$ &  $26$  & 0.7610 & 8 & 0.1747 & 12 & 0.6353 & 19 & 0.0413 \\ \hline
G & LDA & $\sqrt{n/2}$ &  $70$  & 0.7551 & 15 & 0.1740 & 15 & 0.6400 & 9 & 0.0412 \\ \hline
G & LDA & $\#Com$ &  $26$  & 0.7570 & 11 & 0.1742 & 14 & 0.6354 & 18 & 0.0404 \\ \hline
L & AGNES & $\#Com$ &  $6.02 \pm 4.52$  & 0.7602 & 9 & 0.1748 & 10 & 0.6164 & 33 & 0.0395 \\ \hline
G & DIANA & $\sqrt{n/2}$ &  $70$  & 0.7480 & 21 & 0.1723 & 24 & 0.6452 & 7 & 0.0392 \\ \hline
\multicolumn{4}{|c|}{\textbf{M-Prof}}  & 0.7195 & 35 & 0.1724 & 23 & 0.6577 & 1 & 0.0390 \\ \hline
G & KMEANS & $\sqrt{n/2}$ &  $70$  & 0.7484 & 20 & 0.1729 & 20 & 0.6377 & 13 & 0.0387 \\ \hline
L & AGNES & $m*n/t$ &  $15.85 \pm 9.67$  & 0.7476 & 22 & 0.1725 & 22 & 0.6380 & 11 & 0.0385 \\ \hline
G & LDA & $m*n/t$ &  $24$  & 0.7507 & 17 & 0.1727 & 21 & 0.6269 & 28 & 0.0367 \\ \hline
G & SOM-KM & $\#Com$ &  $26$  & 0.7497 & 19 & 0.1730 & 18 & 0.6246 & 31 & 0.0365 \\ \hline
L & PAM & $\sqrt{n/2}$ &  $4.25 \pm 2.60$  & 0.7466 & 24 & 0.1730 & 17 & 0.6277 & 27 & 0.0364 \\ \hline
L & KMEANS & $m*n/t$ &  $15.85 \pm 9.67$  & 0.7421 & 28 & 0.1722 & 26 & 0.6377 & 15 & 0.0363 \\ \hline
G & SOM-KM & $\sqrt{n/2}$ &  $70$  & 0.7441 & 26 & 0.1720 & 27 & 0.6365 & 17 & 0.0361 \\ \hline
G & SOM-KM & $m*n/t$ &  $24$  & 0.7475 & 23 & 0.1730 & 19 & 0.6254 & 30 & 0.0358 \\ \hline
\multicolumn{4}{|c|}{\textbf{I-SubP}} & 0.7505 & 18 & 0.1687 & 33 & 0.6283 & 25 & 0.0353 \\ \hline
L & LDA & $m*n/t$ &  $15.85 \pm 9.67$  & 0.7465 & 25 & 0.1722 & 25 & 0.6280 & 26 & 0.0352 \\ \hline
L & SOM-KM & $m*n/t$ &  $15.85 \pm 9.67$  & 0.7367 & 32 & 0.1711 & 29 & 0.6339 & 21 & 0.0345 \\ \hline
L & DIANA & $m*n/t$ &  $15.85 \pm 9.67$  & 0.7370 & 31 & 0.1702 & 32 & 0.6345 & 20 & 0.0344 \\ \hline
G & KMEANS & $\#Com$ &  $26$  & 0.7429 & 27 & 0.1712 & 28 & 0.6265 & 29 & 0.0341 \\ \hline
L & PAM & $m*n/t$ &  $15.85 \pm 9.67$  & 0.7389 & 30 & 0.1707 & 31 & 0.6290 & 24 & 0.0340 \\ \hline
G & KMEANS & $m*n/t$ &  $24$  & 0.7415 & 29 & 0.1710 & 30 & 0.6218 & 32 & 0.0332 \\ \hline
\multicolumn{4}{|c|}{\textbf{C-SubP}}  & 0.7352 & 33 & 0.1655 & 34 & 0.6108 & 35 & 0.0319 \\ \hline
L & DIANA & $\#Com$ &  $6.02 \pm 4.52$  & 0.7214 & 34 & 0.1652 & 35 & 0.6131 & 34 & 0.0318 \\ \hline
L & KMEANS & $\#Com$ &  $6.02 \pm 4.52$  & 0.7079 & 36 & 0.1622 & 37 & 0.5996 & 36 & 0.0311 \\ \hline
L & PAM & $\#Com$ &  $6.02 \pm 4.52$  & 0.7062 & 37 & 0.1628 & 36 & 0.5861 & 37 & 0.0310 \\ \hline
L & SOM-KM & $\#Com$ &  $6.02 \pm 4.52$  & 0.6913 & 38 & 0.1583 & 38 & 0.5826 & 39 & 0.0305 \\ \hline
L & LDA & $\#Com$ &  $6.02 \pm 4.52$  & 0.6900 & 39 & 0.1576 & 39 & 0.5842 & 38 & 0.0304 \\ \hline
\end{tabular}
\end{small}
\label{tab:ResultadosFiltrado}
\end{table}

\begin{table}[htbp]
\caption{Values of the evaluation metrics for profiles based on clusters and baselines for recommendation (Labels of columns: T = Type of clustering; (L)ocal or (G)lobal; k = method for computing the number of clusters;  $\#Clusters$ = number of clusters; r@10 = recall at 10; P-r = Position in the recall ranking; p@10 = precision at 10; P-p = Position in the precision ranking; ndcg@10 = NDCG at 10; P-ndcg = Position in the NDCG ranking; RRF = Reciprocal Rank Fusion value)}
\begin{small}
\begin{tabular}{|l|l|l|l|c|c|c|c|c|c|c|}
\hline
\multicolumn{1}{|c|}{\textbf{T}} & \multicolumn{1}{c|}{\textbf{Alg.}} & \multicolumn{1}{c|}{\textbf{k}} & \multicolumn{1}{c|}{\textbf{\#Clusters}} & \multicolumn{1}{c|}{\textbf{r@10}} & \multicolumn{1}{c|}{\textbf{P-r}} & \multicolumn{1}{c|}{\textbf{p@10}} & \multicolumn{1}{c|}{\textbf{P-p}} & \multicolumn{1}{c|}{\textbf{ndcg@10}} & \multicolumn{1}{c|}{\textbf{P-ndcg}} & \multicolumn{1}{c|}{\textbf{RRF}} \\ \hline
L & LDA & $m*n/t$ &  $15.85 \pm 9.67$  & 0.6529 & 1 & 0.1486 & 1 & 0.5195 & 2 & 0.0489 \\ \hline
L & KMEANS & $m*n/t$ &  $15.85 \pm 9.67$  & 0.6502 & 2 & 0.1482 & 3 & 0.5214 & 1 & 0.0484 \\ \hline
L & SOM-KM & $m*n/t$ &  $15.85 \pm 9.67$  & 0.6498 & 3 & 0.1482 & 2 & 0.5178 & 4 & 0.0476 \\ \hline
L & PAM & $m*n/t$ &  $15.85 \pm 9.67$  & 0.6481 & 4 & 0.1475 & 4 & 0.5183 & 3 & 0.0471 \\ \hline
G & AGNES & $\sqrt{n/2}$ &  $70$  & 0.6438 & 5 & 0.1465 & 6 & 0.5065 & 9 & 0.0450 \\ \hline
G & DIANA & $\sqrt{n/2}$ &  $70$  & 0.6408 & 7 & 0.1459 & 8 & 0.5163 & 5 & 0.0450 \\ \hline
G & SOM-KM & $\sqrt{n/2}$ &  $70$  & 0.6437 & 6 & 0.1470 & 5 & 0.5023 & 10 & 0.0448 \\ \hline
G & LDA & $\sqrt{n/2}$ &  $70$  & 0.6398 & 8 & 0.1459 & 7 & 0.5022 & 11 & 0.0437 \\ \hline
L & DIANA & $m*n/t$ &  $15.85 \pm 9.67$  & 0.6385 & 9 & 0.1453 & 11 & 0.5080 & 8 & 0.0433 \\ \hline
G & DIANA & $\#Com$ &  $26$  & 0.6357 & 13 & 0.1445 & 17 & 0.5113 & 6 & 0.0418 \\ \hline
G & DIANA & $m*n/t$ &  $24$  & 0.6364 & 11 & 0.1445 & 18 & 0.5107 & 7 & 0.0418 \\ \hline
G & KMEANS & $\sqrt{n/2}$ &  $70$  & 0.6380 & 10 & 0.1452 & 12 & 0.4961 & 16 & 0.0413 \\ \hline
L & KMEANS & $\sqrt{n/2}$ &  $4.25 \pm 2.60$  & 0.6350 & 15 & 0.1450 & 13 & 0.4976 & 12 & 0.0409 \\ \hline
G & LDA & $m*n/t$ &  $24$  & 0.6342 & 16 & 0.1446 & 16 & 0.4962 & 13 & 0.0400 \\ \hline
L & LDA & $\#Com$ &  $6.02 \pm 4.52$  & 0.6352 & 14 & 0.1457 & 9 & 0.4717 & 29 & 0.0392 \\ \hline
L & AGNES & $m*n/t$ &  $15.85 \pm 9.67$  & 0.6322 & 17 & 0.1443 & 19 & 0.4962 & 14 & 0.0392 \\ \hline
L & LDA & $\sqrt{n/2}$ &  $4.25 \pm 2.60$  & 0.6320 & 18 & 0.1442 & 20 & 0.4961 & 15 & 0.0387 \\ \hline
G & LDA & $\#Com$ &  $26$  & 0.6316 & 19 & 0.1440 & 22 & 0.4951 & 17 & 0.0378 \\ \hline
L & SOM-KM & $\sqrt{n/2}$ &  $4.25 \pm 2.60$  & 0.6313 & 20 & 0.1441 & 21 & 0.4950 & 18 & 0.0377 \\ \hline
L & SOM-KM & $\#Com$ &  $6.02 \pm 4.52$  & 0.6295 & 21 & 0.1449 & 14 & 0.4665 & 32 & 0.0367 \\ \hline
L & KMEANS & $\#Com$ &  $6.02 \pm 4.52$  & 0.6284 & 22 & 0.1447 & 15 & 0.4685 & 31 & 0.0365 \\ \hline
 \multicolumn{4}{|c|}{\textbf{C-SubP}}  & 0.6048 & 38 & 0.1457 & 10 & 0.4795 & 25 & 0.0363 \\ \hline
G & SOM-KM & $m*n/t$ &  $24$  & 0.6281 & 23 & 0.1435 & 25 & 0.4807 & 22 & 0.0360 \\ \hline
L & DIANA & $\sqrt{n/2}$ &  $4.25 \pm 2.60$  & 0.6262 & 26 & 0.1430 & 27 & 0.4892 & 19 & 0.0358 \\ \hline
G & KMEANS & $m*n/t$ &  $24$  & 0.6278 & 24 & 0.1437 & 24 & 0.4796 & 24 & 0.0357 \\ \hline
G & SOM-KM & $\#Com$ &  $26$  & 0.6254 & 27 & 0.1432 & 26 & 0.4804 & 23 & 0.0352 \\ \hline
L & PAM & $\sqrt{n/2}$ &  $4.25 \pm 2.60$  & 0.6247 & 29 & 0.1426 & 29 & 0.4880 & 20 & 0.0350 \\ \hline
L & DIANA & $\#Com$ &  $6.02 \pm 4.52$  & 0.6262 & 25 & 0.1440 & 23 & 0.4687 & 30 & 0.0349 \\ \hline
 \multicolumn{4}{|c|}{\textbf{M-Prof}}  & 0.6358 & 12 & 0.1357 & 38 & 0.4546 & 35 & 0.0346 \\ \hline
G & KMEANS & $\#Com$ &  $26$  & 0.6248 & 28 & 0.1429 & 28 & 0.4791 & 26 & 0.0344 \\ \hline
L & AGNES & $\sqrt{n/2}$ &  $4.25 \pm 2.60$  & 0.6215 & 30 & 0.1422 & 32 & 0.4747 & 27 & 0.0335 \\ \hline
G & PAM & $\sqrt{n/2}$ &  $70$  & 0.6151 & 33 & 0.1398 & 33 & 0.4721 & 28 & 0.0329 \\ \hline
 \multicolumn{4}{|c|}{\textbf{I-SubP}} & 0.5959 & 39 & 0.1355 & 39 & 0.4868 & 21 & 0.0325 \\ \hline
L & PAM & $\#Com$ &  $6.02 \pm 4.52$  & 0.6173 & 31 & 0.1423 & 31 & 0.4537 & 36 & 0.0324 \\ \hline
L & AGNES & $\#Com$ &  $6.02 \pm 4.52$  & 0.6172 & 32 & 0.1423 & 30 & 0.4440 & 39 & 0.0321 \\ \hline
G & AGNES & $m*n/t$ &  $24$  & 0.6123 & 34 & 0.1392 & 34 & 0.4548 & 34 & 0.0319 \\ \hline
G & AGNES & $\#Com$ &  $26$  & 0.6119 & 35 & 0.1389 & 35 & 0.4548 & 33 & 0.0318 \\ \hline
G & PAM & $m*n/t$ &  $24$  & 0.6097 & 36 & 0.1387 & 36 & 0.4520 & 37 & 0.0311 \\ \hline
G & PAM & $\#Com$ &  $26$  & 0.6089 & 37 & 0.1384 & 37 & 0.4502 & 38 & 0.0308 \\ \hline
\end{tabular}
\end{small}
\label{tab:ResultadosRecomendacion}
\end{table}

First, we shall focus on the performance of baseline methods in both recommendation and filtering tasks, taking into account the aggregated ranking of evaluation measures. The first conclusion is that all of them are placed in the lower half of both tables. This means that there is a good number of clustering algorithms that outperform them. In terms of performance and in the context of filtering, it is noticeable that monolithic profiles and intervention-based subprofiles are better than committee-based ones. Focusing on the recommendation problem, the best baseline is committee-based subprofiles and the worst is intervention-based subprofiles.

Considering the filtering problem, we consider recall to be the most interesting metric because the system should retrieve as many relevant MPs as possible, i.e. identify the highest number of MPs interested in the document to be filtered. If we look at the expert finding problem, we believe NDCG to be a more valuable metric than recall because in this case, we are not only interested in retrieving the largest number of relevant MPs but also we want them to be ranked highest. In both cases, and focusing on the corresponding metric, we notice that monolithic profiles perform worst and the intervention-based subprofiles the best, while committee-based subprofiles are placed between them.

Observing the results contained in these tables we could say that there is quite a good number of clustering algorithms that perform better than the baseline profiles both for recommendation and filtering tasks. In both cases, more than half of the clustering methods outperform the baselines C-SubP and M-Prof (the given political clusters and the option of no clustering at all, respectively), considering the final combined ranking. This number increases to two thirds when we focus on recall for filtering and NDCG for recommendation.

Table \ref{tab:porcentajesMejora} shows the improvement percentage of the best clustering algorithms for filtering and recommendation, considering recall and NDCG, respectively, with respect to the baselines. These percentages are moderate but reflect the fact that clustering is a good alternative for capturing the underlying topics and creating subprofiles. We should highlight that the greatest improvement percentages are achieved for M-Prof, which is good news because it supports the fact that the use of subprofiles by clustering initiatives is better than using a single profile. These percentages are lower when compared with C-SubP but are still relevant, and this supports our hypothesis that political divisions in certain cases may well be somewhat artificial. It is also worth mentioning that the differences between the top clustering methods and baselines are always statistically significant (using a t-test) as occurs with most of the clustering algorithms placed above the baselines.

\begin{table}[htbp]
\caption{Improvement percentages of the best clustering methods for the baselines. $*$ means a statistically significant difference}
\begin{center}
\begin{tabular}{|c|c|c|}
\hline
 &  \textbf{Filtering – Recall} & \textbf{Recommendation - NDCG} \\ \hline
 & \textbf{Global AGNES $\sqrt{n/2}$} & \textbf{Local LDA m*n/t} \\ \hline
\textbf{M-Prof} & 7.35 \% *& 14.27 \% *\\ \hline
\textbf{C-SubP} & 5.06 \% *& 8.33 \% *\\ \hline
\textbf{I-SubP} & 2.91 \% *& 6.72 \% *\\ \hline
\end{tabular}
\end{center}
\label{tab:porcentajesMejora}
\end{table}

The general conclusion of this analysis, and to answer the second research question proposed in this section, is that clustering-based subprofiles are a good option for filtering and recommendation tasks since they perform better than baseline approaches. In our opinion, it is much better and more natural that the fixed committee groups, which are constructed from committees that have in turn been created for political reasons, because they are able to represent the topics that a user is interested in more precisely. Clustering enables the creation of different clusters for topics from the same committee or the combination of facets that might have been artificially separated into two different committees, and it is of course a much better approach than to create a single profile where all the topics are jumbled up.

\subsubsection{RQ3: Is there any difference between building the clusters locally or globally?}

The next step is to determine whether the best approach is global or local clustering. In the case of filtering new documents, and focusing both on recall and the ranking of the combined metrics, it is remarkable how global clustering is superior to the local alternative: most top clustering algorithms used to build the subprofiles use a global approach, while the local grouping techniques perform worse. This clear distinction in terms of performance is not so evident when we focus on the politician recommendation problem (NDCG and combined metrics). In this case, the global and local clustering algorithms are more mixed throughout the ranking, but it is true that the best clustering algorithms positively employ the local approach. To support numerically this conclusion, we have computed the average rank where the methods using global and local clustering appear. For the filtering problem these average ranks are $15.28$ for global and $23.67$ for local clustering. However, for the recommendation problem, these averages are nearer, $20.94$ for global and $17.22$ for local clustering.

There is one possible explanation for this behaviour: the local approach forces the interventions of a given MP to be distributed among exactly $k$ clusters and this may be a more artificial division in some cases. On the other hand, in a global approach, these MP's documents are probably not assigned to all these $k$ clusters and they could therefore be divided into more cohesive and natural groups. This means that profile sizes in the local approach could be smaller than those built with the global one. In a filtering setting, since the query is the full text of an initiative, it is quite a large query in comparison with the recommendation problem where the query  is basically a paragraph with a few lines. Our conjecture is that large queries perform better with large subprofiles as occurs with the global approach in the filtering context. When the query is much shorter (recommendation problem), subprofile lengths are not so important and so global and local clustering are much more mixed. In terms of the answer to $RQ3$, we would say that the global approach is more interesting for filtering and local somewhat better for recommendation.

\subsubsection{RQ4: Is the number of clusters relevant for the recommendation quality?}

As we have already mentioned, selecting the number of clusters is the main problem that computer or data scientists must face in the context of clustering. In our case, we have tried three different methods for computing such a value although we do not intend to exhaustively try many methods and find the best but simply to check the sensibility of different values for this parameter. In addition to the well-known $\sqrt{n/2}$ and $m*n/t$, we have also used the number of committees as a kind of baseline for $k$. In global clustering, the values of $\#Com$ and $m*n/t$ are very close ($26$ and $24$, respectively) and so their results are quite similar, independently of the type of problem and clustering algorithm.

For the filtering problem and with the global strategy, all of these methods combined with any $k$, including $k=\#Com$, are better than C-SubP for both the recall metric and the combination of metrics and most are better than I-SubP and M-Prof. The best value for $k$ in absolute numbers is $\sqrt{n/2}$. We believe that there is more room for including new subtopics when $k=70$ than with $26$, where these are grouped together in the same clusters, and so this is a more robust value for every clustering algorithm. For the local mode, meanwhile, it is noticeable how the performance of most clustering algorithms is really bad when $k=\#Com$ is applied and even worse than C-SubP. On the other hand, $\sqrt{n/2}$ is again the method that behaves best. The reason for this may be that in the local case $\sqrt{n/2}$ is the method that generates the lowest mean number of clusters ($4.25$) and so subprofile sizes are larger and this tendency is positive in the filtering problem.

For recommendation and global clustering, $\sqrt{n/2}$ performs best and is very robust across clustering methods. The performance of $\#Com$ and $n*m/t$ clearly depend on the algorithm but it is generally much worse (for NDCG and the combined ranking). Focusing on local clustering, it seems that $\#Com$ and $\sqrt{n/2}$ do not provide enough space to include the different topics that MPs deal with and more groups are required and so $n*m/t$ is the best alternative (it performs best and every clustering method achieves the best values with it). Any algorithm combined with $n*m/t$ does in fact outperform C-SubP and the other baselines.

In terms of individual algorithms, AGNES is very robust independently of the $k$ selected for filtering in local and global approaches, and for the recommendation problem, the performance of this clustering clearly varies according to it. For the remaining algorithms, it is not possible to draw such an obvious conclusion since performance varies according to the number of clusters used, the type of clustering and the problem at hand.

By way of conclusion and to answer RQ4, we would say that selecting a good value for the $k$ parameter is important for good performance in the filtering and recommendation problems. We should also mention that we have found $k$ values that outperform the number of committees. This means that we cannot restrict MPs only to the committees where they intervene.

\subsubsection{RQ5: What are the best clustering algorithms for these tasks?}

As previous step to answer this question, we have carried out two ANOVA variance tests, with $\alpha=0.05$, one for the results of recall@10 for filtering and the other for those of ndcg@10 for recommendation, for all the combinations. The conclusion is that there are significant differences among them (p-values of $4,9309E-37$ and $1.1842E-24$, respectively). Therefore, it is important to make a good decision about the clustering method in combination with the local or global approaches and the number of clusters in order to get a good performance in these problems. 

The clustering techniques that perform best vary according to the problem at hand. For the filtering problem, hierarchical clustering techniques work quite well and AGNES, in particular, is the best algorithm in its global clustering version in terms of the four evaluation metrics used. For the expert finding problem, although good results are also obtained by AGNES, the best approaches are Local LDA in most of the metrics, followed by the centroid-based algorithm Local K-MEANS, the SOM-KM approach and PAM. In this case, we could observe how recommendation could be performed with quality using a wide variety of clustering techniques. We have performed another ANOVA test with the top 5 combinations of Tables \ref{tab:ResultadosFiltrado} and \ref{tab:ResultadosRecomendacion}, again with the recall@10 and ndcg@10 values, respectively, and the result is that there are no significant differences among them (p-values of $0.8154$ and $0.9691$, respectively). This means that any of them could be selected for these tasks with high confidence of doing a good job. However, we should mention that the performance of the clustering algorithms clearly depends on the value of the $k$ parameter, as we have discussed in the previous section.

\begin{figure}[tb]
\centering{
\includegraphics[width=1.0\textwidth]{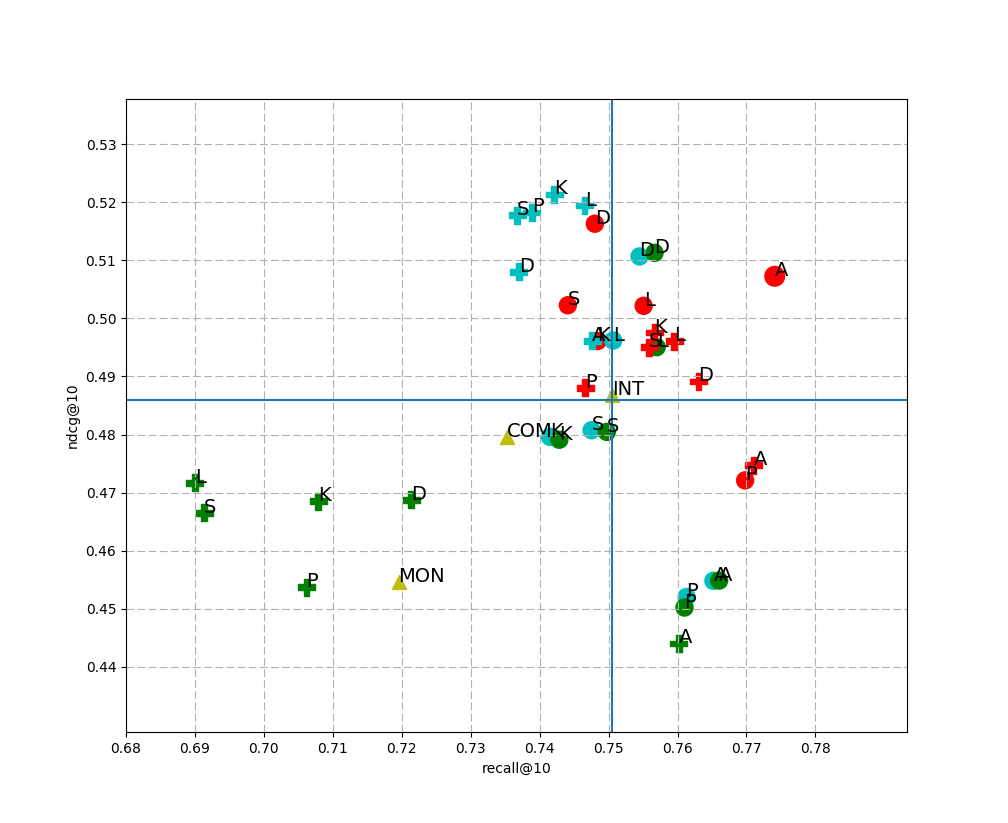}
\caption{recall@10 vs ndcg@10 of all the clustering combinations plus the baselines}
\label{fig:recallVSndcg}
}
\end{figure}

Finally, we have plotted the recall@10 (for filtering) and ndcg@10 (for recommendation) values of all the clustering combinations in a graph (Figure \ref{fig:recallVSndcg}) in order to graphically discover the combinations with better performance in both tasks. In this plot, we have used the different shapes to represent the global (circle) or local (filled plus sign) clustering. Also, different colors to represent the parameter $k$, i.e., the number of clusters, being $\#Com$, $\sqrt{n/2}$ and $m*n/t$ represented by green, red and blue, respectively. Finally, the type of cluster is represented by the first letter of the name of the technique: Kmeans, Lda, Diana, Agnes, Som, Pam. We have also included the three baselines (MONolithic, COMmittee-based and INTervention-based), represented with a triangle in the graph.

From this graph we can see that using $\#Com$ as the number of clusters is not a good alternative for  filtering (where the worse results has been obtained) and also for recommending. This can be considered as an evidence that the number of committees in the parliament does not match properly with the topics discussed, and therefore to obtain the best results is not essential to know a priori this information. This   is an interesting result, since our approach could be extended to other problems where such information is not available.

Also, if we want to find one strategy which fits both filtering and recommendation tasks  we propose to focus on those methods which are better than the best baseline (intervention-based) for both problems, i.e., those situated in the top-right area of the graph (delimited by solid lines). In this case, we can see that using a global hierarchical or LDA as clustering algorithm with $\sqrt{n/2}$ as the number of clusters are reasonable alternatives, obtaining the best results with AGNES-Global-$\sqrt{n/2}$. Nevertheless, if we focus only in the problem of recommending,  this algorithm seems to be very dependent on the value of the other parameters, being necessary a proper estimation of the parameters. Therefore, if we are looking for an good cluster strategy, suitable for both problems, the most stable algorithms seem to be both LDA and DIANA, particularly using global clustering.


\section{Related work} \label{section:related}

In this section, we shall present relevant published work on the general subject of compound profiles. Although some of these publications do not use clustering, we thought that it would be interesting to include them in this review since we represent multi-faceted profiles in our work and most authors agree that users would benefit from a richer representation. In order to summarise the papers, we have included Table \ref{tab:CompoundProfiles} in the section with the following main features: profile purpose (Purpose); information source used (Inf. Source); clustering algorithm used (Clust. Alg.), if necessary; entities considered (Entity cons.) and their features (Features); and finally, the type of compound profile obtained (Type of profile). We have also included our approach in the bottom row of the table.

A first group of more closely related papers deals with a structured way of representing profiles based on different information sources. In the context of expert finding, \citep{PL15} creates a structured profile with experts' personal data, expertise and interests, represented by a tree with three sibling nodes, typically containing the terms from each source. A second case of this style is \citep{HMOS12}, where based on Twitter topical lists, two subprofiles are built for a user: one comprising the tags assigned to the lists to which they belong and the other with the tags from the lists for the user's friends. A third example is presented in \citep{ZCM02}, where user evidence is represented by different types of objects (Web pages, users, items, queries, etc.), which are clustered in a multi-layered graph, creating cluster connections by applying mutual reinforcement. Finally, in \citep{NMSLG13}, the authors, with the aim of recommending social items, create three subprofiles using as sources the weighted keywords extracted from the user's social items, tags associated to these keywords, and new terms connected by underlying concepts.


Other related papers present approaches where the information coming from one source (typically documents) is organized into different profiles. One first case designed for news recommendation (\citep{GFSC14}) proposes a method for representing two faceted-profiles: a long-term subprofile comprising terms and categories from the history of relevant documents; a short-term one, with the same information but created after the first subprofile has been built. A second example is \citep{BT17}, where the profile comprises two subprofiles: the list of terms extracted from positively judged documents, enriched by the terms belonging to the cluster to which the user belongs (after applying K-means to every user) and enriched by Wordnet hypernyms, and the terms from negatively judged documents. 

Another type of structured profile is the one based on hierarchies. The paper \citep{RBN14} builds expertise profiles comprising a series of time-based hierarchical profiles, where the nodes are weighted topics. \citep{CCS01} presents a personalization system based on keeping a hierarchy of the user's interests (personal view) from visited web pages. In both cases, the hierarchies are given not learned.

Considering a structure where a profile comprises a list of categories/topics/concepts which are not interlinked, and each is represented by a set of keywords, we can cite the recommender {\em Syskill \& Webert} \citep{PMB96}, the personalizer {\em Alipes} \citep{WYNYZY99} and the recommender {\em Webmate} \citep{CS98}. In these first two cases, the category list is given to the system in one way or another, and not learned automatically as it is in the third system which uses a clustering algorithm. Another example in this category is presented in \citep{Koo05}, where category-based subprofiles are created not with documents but with terms from the formulated queries (clustering queries). 

The following papers offer a similar profile structure but are automatically learned by clustering: \citep{SH01} applies a local incremental clustering to generate topics from clusters for search personalization; {\em Web Personae} \citep{MKS02} uses a hierarchical clustering on the user's visited set of web pages (local construction) for the same personalization purposes. 

In \citep{AIOS14}, once the terms have been extracted from the information sources, the induced bisecting K-means algorithm is applied to group these terms into semantically related concepts, representing each user (scholars, in this case) with a set of research areas (groups) and characterized by a set of terms. A similar approach is presented in \citep{YGS09}, where using a document clustering technique based on community-discovery methods, the authors create groups of tags to represent the users. While this is a local tag-based clustering, we also consider global clustering but with terms as features. \citep{PCBC07} essentially presents the same idea as \citep{YGS09}, grouping similar documents, but the main difference is that the first does not explicitly represent user interests but uses the structure of clusters to directly recommend scientific articles.


\begin{landscape}
\begin{table}[tbp]
\begin{center}
\begin{scriptsize}
\begin{tabular}{|m{30pt}|m{105pt}|m{70pt}|m{70pt}|m{60pt}|m{50pt}|m{170pt}|}
\hline
\bf{Reference} & \bf{Purpose} & \bf{Inf. Source} & \bf{Clust. Alg.} & \bf{Entity Cons.} & \bf{Features} & \bf{Type of profile} \\ \hline \hline
\citep{PL15} & Expert finding & Heterogeneous documents & -- & Documents & Keywords & Tree with nodes containing keywords \\ \hline
\citep{HMOS12} & User modeling & Lists in Twitter & -- & -- & Tags assigned to lists & Intentional (weighted tags in lists the user follows) and Extensional profiles (weighted tags in lists the friends follows) \\ \hline
 \citep{ZCM02} & User modeling & Web heterogeneous objects & Probabilistic clustering & Web Objects & Keywords & Multi-layered graph with nodes in different layers representing different type of objects \\ \hline
 \citep{NMSLG13} &Content-based recommendation  & Social items from Facebook \& Instagram & -- & -- & Keywords \& Concepts & Wikipedia Concepts \& Extended keywords \\ \hline \hline

  \citep{BT17} & Content-based recommendation & News & Variation of K-Means & Users & Keywords & Terms in positive documents (and in clusters they belong to), in negative documents, plus WordNet hypernyms \\ \hline
  \citep{GFSC14} & Content-based recommendation & News & -- & News & Keywords & Keywords in observed News, keywords in concepts \\ \hline \hline

   \citep{RBN14} & IR Personalization & Web pages & -- & User's interest & Keywords & Hierarchy of web pages \\ \hline
   \citep{CCS01} & IR Personalization & Web pages & -- & Web pages & Keywords & Hierarchy of the user's interests \\ \hline \hline

 \citep{PMB96} & Content-based recommendation & Web pages \& their categories & -- & -- &Keywords \& Categories & List of categories and the associated terms \\ \hline
  \citep{WYNYZY99} & News personalization & News & -- & -- & Keywords and categories & List of categories comprising three lists of keywords, respectively \\ \hline
  \citep{CS98} & Content-based recommendation & Web pages & -- & Keywords & -- & List of categories comprising keywords \\ \hline
\citep{Koo05} & Collaborative tagging & Documents & Community Discovery-based & Documents & Keywords & Subprofiles comprising keywords \\ \hline \hline

 \citep{SH01} & IR Personalization & Web pages & Incremental clustering & Web pages & Keywords &  List of topics \\ \hline
 \citep{MKS02} & IR Personalization & Web pages & Hierarchical clustering & Web pages & Keywords & List of cluster centroids \\ \hline \hline

 \citep{PCBC07} & Content-based recommendation & Scientific articles & -- & Scientific articles & Keywords & Clusters of articles \\ \hline
 \citep{YGS09} & Collaborative tagging & Documents & Community discovery technique & Documents & Keywords & Subprofiles comprising tags (extracted from clustered documents) \\
 \citep{AIOS14} & Expert finding & Heterogeneous sources & Bisecting K-Means & Different sources & Keywords & Subprofiles of research areas \\ \hline \hline

de Campos et al. (2018) & Content-based recommendation \& filtering & Parliament initiatives & AGNES, DIANA, LDA, K-Means, PAM, SOM & Initiatives & Keywords & List of subprofiles, containing weighted keywords \\ \hline
\label{tab:CompoundProfiles}
\end{tabular}
\end{scriptsize}
\caption{Summary of the related works on compound profiles}
\end{center}
\end{table}
\end{landscape}

\subsection{Differences of our approach with the related work}

The first four approaches presented in this review (\citep{PL15,HMOS12,NMSLG13,ZCM02}) in the context of a structured way of representing profiles based on different information sources, differ from our proposal in that they consider various sources of information to build the profiles whereas we only take into account one type and the profile structure is also relatively complex to support this diversity. Another difference is that with the exception of \citep{NMSLG13}, none of the papers is interested in capturing and representing underlying topics as we are. This referenced paper uses concepts that are extracted from an external source and not automatically learnt as we implicitly do. Finally, none of them use clustering to build profiles with the exception of \citep{ZCM02}.

In the case where information coming from one source (typically documents) is organized into different profiles \citep{GFSC14}, in our proposal, we do not consider positive and negative documents. In the second reference, \citep{BT17}, the authors apply global clustering at the user level (users are the instances and the terms, the attributes) while our clustering is carried out at the document level and locally (only for the active user).

The main difference with the approaches presented in \citep{RBN14,CCS01} is obvious as they do not use clustering and we do not use concept hierarchies to represent the profiles. In our case, the concepts are not interlinked.

When focusing on profiles composed of a list of categories, topics or concepts \citep{PMB96,WYNYZY99,CS98,Koo05}, although the profile structure is very similar to the one presented in \citep{Koo05}, the main difference is that we use proper clustering algorithms to automatically create the categories (the clusters). These approaches construct the profiles locally. Our proposal also considers the global information of all the users. Additionally, we use the document terms as features whereas in \citep{Koo05} query terms are used.

With respect to papers \citep{SH01} and \citep{MKS02}, which build a similar profile structured by means of clustering, once again, the main difference with our approach is the locality of profile construction. The use of profiles is also another difference since in these papers, profiles are considered to be personalization tools, whereas in our case, they are used for content-based recommendation. Finally, a third difference is profile selection because these examples use the most relevant profile (only one) while our proposal combines all of these results to find the user to recommend to.

Considering the papers \citep{AIOS14,YGS09,PCBC07}, the main difference is that, in our case, the clusters contain documents and we also consider global clustering but with terms as features. 

In addition to the differences described in this section, we should mention that our experiments have tested the suitability of different clustering algorithms and different methods to decide on the number of clusters to be used. It is difficult to find any specific published reference as to how the number of clusters should be determined.

\section{Conclusions and Further Research} \label{section:conclusions}

In this paper, we have presented a proposal based on text clustering to automatically build compound profiles of experts to properly reflect the topics in which they are usually interested. Two different but highly related application domains have been considered, namely filtering and expert recommendation. In the first case the task is to decide which experts would be interested in receiving a new document, according to their interests and expertise. In the second case the decision to be made is which experts are more appropriate to satisfy an information need expressed by a user. The specific setting where we have experimentally tested our proposals is political, where the experts considered are Members of Parliament and the information source used to build the profiles are the transcriptions of their interventions when discussing initiatives within the parliamentary debates.

Although these two problems, filtering and recommendation can be formulated in a unified way (given a query, either a document to be filtered or an information need to be satisfied, return a ranked list of experts which are either interested in the document or able to satisfy the information need) and both can be managed using a similar approach, our experimental results suggest that there are some important differences between them. These differences determine that the more appropriate tools for solving these problems within our formulation (type of clustering, local or global, type of clustering algorithm and selection of the number of clusters) are different.

We have proposed two clustering alternatives: a local method and a global method. The local method separately clusters the documents of each expert (i.e. the interventions of each MP), whereas the global method performs a single clustering of the documents of all the experts. In any case we have tested clustering algorithms of very different nature (hierarchical agglomerative and divisive, centroid-based, generative statistical model-based, neural network-based), as well as different techniques to select the number of clusters. Three different baselines have been considered: two extreme cases, a single (monolithic) profile for each MP and one subprofile for each MP intervention, and an intermediate situation where the subprofiles of an MP are not learned through clustering but each subprofile is extracted from the interventions of the MP in each of the different committees where she participates.

The main conclusions extracted from our experimental results are the following:
\begin{itemize}
\item Clustering is generally a good option to discover groups of documents dealing with different topics of interest, even improving situations where these groups are given explicitly and externally.
\item Many of the alternatives based on clustering outperform all the three baseline methods for both filtering and recommendation, the differences in performance being statistically significant.
\item Concerning the type of clustering, for the filtering problem the choice is clear: global clustering is preferable. However, for the recommendation problem the situation is not so clear, although the four top performing alternatives use local clustering. This different behavior seems to be related to the sizes of the clusters generated by each approach and the fact that in the filtering problem the size of the ``query'' (the complete document) is normally larger than in the recommendation problem (where the ``query'' is the information need of a user).
\item If we focus on the selection of the number of clusters, we again find differences between filtering and recommendation: for filtering the best method to select the number of clusters is $\sqrt{n/2}$. For recommendation, however, $n*m/t$ performs best.
\item Concerning the specific clustering algorithms being used, hierarchical methods (in particular the agglomerative one) work quite well for the filtering problem, whereas LDA, centroid-based methods and SOM are preferable for the recommendation problem. However, it seems that the decision about which clustering algorithm to use is not critical, because we have not found statistically significant differences among the five best performing methods within each problem.
\end{itemize}

By way of future research, we plan to tackle the problem of how recommendation and filtering problems would be affected when experts are represented by temporary profiles. In this case, short and long profiles would be built for them using clustering techniques. Another line that we would like to explore is the potential capacity of LDA and other topic models \citep{JLZ18,LL12,MGB17} for creating good (sub)profiles but by exploiting the semantic perspective in which these algorithms specialize.

\section*{Acknowledgements}
This work has been funded by the Spanish Ministerio de Econom\'ia y Competitividad under project TIN2016-77902-C3-2-P, and the European Regional Development Fund (ERDF-FEDER).

\end{document}